\documentclass[11pt,a4paper]{article}
\usepackage{jcappub}
\newcommand{\kk}{\mathbf{k}}
\newcommand{\pp}{\mathbf{p}}

\newcommand{\qq}{\mathbf{q}}
\newcommand{\xx}{\mbox{\boldmath $x$}}

\title{Consistency Relation for multifield inflation scenario with all loop contributions}
\author{Naonori S. Sugiyama}
\affiliation{Astronomical Institute, \\
Graduate school of Science, Tohoku University, Sendai 980-8578, Japan}

\emailAdd{sugiyama@astr.tohoku.ac.jp}

\abstract{
The consistency relation between non-linear parameters $f_{NL}$ and $\tau_{NL}$ 
characterizing Non-Gaussianity generated in the super horizon scale 
have been emerged as a useful tool to rule out a large class of inflationary models all at once.
In our previous work, 
we extended the Suyama-Yamaguchi inequality up to 1-loop corrections.
In this paper, we further extend the inequality up to {\it all} loop corrections, 
and found that it takes the same expression with the original Suyama-Yamaguchi inequality, $\tau_{NL} \geq \left( \frac{6}{5}f_{NL} \right)^2$,
where the equality is not satisfied in the case of single field models any more.
}

\arxivnumber{1201.4048}

\begin{document}
\maketitle

\section{Introduction}
The inflationary scenario is now regarded as an indispensable ingredient in modern cosmology\cite{Sato:1981,Guth:1981,Linde:1982,Linde:1983,Albrecht/Steinhardt:1982}.
One of the most important predictions is the explanation of the origin of the primordial curvature perturbation as vacuum quantum fluctuations.
The predicted curvature perturbation has definite statistical properties, such as  the adiabaticity, nearly scale-invariance and nearly Gaussianity 
\cite{Mukhanov/Chibisov:1981,Guth/Pi:1982,Hawking:1982,Starobinsky:1982,Bardeen/Turner/Steinhardt:1983}. It turns out that these predictions are consistent with the current observation of cosmic microwave background (CMB) 
\cite{Komatsu/etal:2009,Komatsu/etal:2011}.
Nevertheless, we still do not know the detailed model for the inflationary scenario.

In this respect the possible observation of {\it Non-Gaussianity} in the primordial curvature perturbation in CMB temperature fluctuation has attracted much attention recently  because it has a potentiality to discriminate various models of the inflationary scenario. 
To extract the useful information from the observational results it is necessary to have definite theoretical predictions in the nature of the Non-Gaussianity. 
This is our purpose in this paper. 

Here we shall focus on the so-called ``local form'' {\it bispectrum}
\cite{Gangui/etal:1994,Verde/etal:2000,Komatsu/Spergel:2001}
and {\it trispectrum}\cite{Boubekeur/Lyth:2005} in the primordial curvature perturbation. 
In general the bispectrum and trispectrum are defined as follows.  
\begin{equation}
	\langle \prod_{i=1}^{3}\zeta(\kk_i)\rangle 
   = (2\pi)^3 \delta^{3}(\sum_i\kk_i) B_\zeta(k_1,k_2,k_3)  ,
		\label{}
\end{equation}
\begin{equation}
	\langle \prod_{i=1}^{4}\zeta(\kk_i)\rangle 
   =(2\pi)^3 \delta^{3}(\sum_i\kk_i) T_\zeta(k_1,k_2,k_3,k_4)	.
		\label{}
\end{equation}
The local form of them is given by the product of the power spectrum.
The local form bispectrum and trispectrum are characterized by three parameters $f_{\rm NL}$, $\tau_{NL}$ and $g_{NL}$ as follows.
\begin{equation}
 B_\zeta=\frac65f_{\rm
  NL}\left[P_\zeta(k_1)P_\zeta(k_2)+(\mbox{2 perm}.)\right],
\end{equation} 
\begin{align}
  T_{\zeta} =&  \tau_{\rm NL}[P_{\zeta}(|\kk_1 + \kk_3|) P_{\zeta}(k_3)
			P_{\zeta}(k_4) + (\mbox{11 perm.})] \notag \\
			& +  \frac{54}{25}g_{\rm NL}[P_{\zeta}(k_1) P_{\zeta}(k_2) P_{\zeta}(k_3) + (\mbox{3 perm.})].
\end{align}
There is an infinite number of higher moments defined as above. 
In this paper we focus only on $f_{NL}$ and $\tau_{NL}$. 
The current best limit is $f_{NL} = 32 \pm 21$ (68\% CL;\cite{Komatsu/etal:2009,Komatsu/etal:2011}).
The Planck satellite is expected to reduce the error bar by a factor of four \cite{Komatsu/Spergel:2001}.

Recently, various consistency relations satisfied  by non-linear parameters 
have emerged as a useful test of a large class of inflationary models.
In the single field case, it is shown that a convincing detection of $f_{NL} \gg 1$ or $\tau \gg 1$ rules out {\it all} single-field inflation models
\cite{Creminelli/Zaldarriaga:2004,Maldacena:2003,Seery/Lidsey:2005,
Chen/etal:2007,Cheung/etal:2008,Ganc/Komatsu:2010,Renaux-petel:2010,Acquaviva/etal:2002,Sandipan:prep}.  
In the multi field case, Suyama and Yamaguchi proved the following inequality between $f_{NL}$ and $\tau_{NL}$ \cite{Suyama/Yamaguchi:2008} which we call SY inequality.
\begin{equation}
\tau_{NL} \geq \left( \frac{6}{5} f_{NL} \right)^2 ,
\end{equation}
This inequality is derived using the $\delta N$ formalism
\cite{Starobinsky:1982,Starobinsky:1986,Sasaki/Stewart:1996,Lyth/Rodriguez:2005,Valenzuela_Taledo/etal:2011}
in which the curvature perturbation is described by the departure of the $e$-folding 
and expanded by the scalar fields at near horizon crossing. 
In the original SY inequality, 
it is further assumed that non-linear parameters $f_{NL}$ and $\tau_{NL}$ are
generated only on super horizon scales 
and the $\delta N$ expansion is truncated up to the second order, that corresponds to the tree level in a diagrammatical method. 
As the extension of the SY inequality, Suyama et al. \cite{Suyama/etal} pointed out that the truncated $\delta N$ expansion contains a part of 1-loop corrections and the original form of the SY inequality is still satisfied including such 1-loop corrections.  In our previous work, we truncated the $\delta N$ expansion up to the forth order which corresponds to take into account all of the 1-loop corrections \cite{N.S.Sugiyama/Komatsu/etal:2011}.
On the other hand, 
as pointed out in~\cite{K.M.Smith/M.LoVerde/etal:2011}, 
the SY inequality is automatically satisfied by their definition of $f_{\rm NL}$ and $\tau_{NL}$ 
regardless of the details of the inflationary model in infinitely large universe,
and it is also suggested that the definition itself leads to the weaker condition for a finite observable universe
\footnote {The Eq. (9) in \cite{K.M.Smith/M.LoVerde/etal:2011} is given by
\begin{equation}
		 \tau_{NL} \geq \left( \frac{6}{5} f_{NL} \right)^2 - \frac{1}{2P_{\zeta}V}.
		\label{weaker}
\end{equation}
The second term of the Right Hand Side means the correction of a finite volume of the observable universe.
This term vanish in the infinite universe, but  always become negative in a finite observable universe.
Therefore, this inequality is weaker than the SY inequality.
However, it is not trivial what the specific difference between Eq.~(\ref{weaker}) and the SY inequality shown form 
the $\delta N$ formalism is}.
Thus there remains still a possibility that the SY inequality is violated in a finite observable universe,
and it is interesting to investigate the generality of the SY inequality from a physical point of view.
This is important because future CMB observation such as the Planck satellite might detect these parameters in a good accuracy
and give us a useful information on the inequality. 

To study the generality of the SY type inequality,
we need to extend the SY inequality up to {\it all} order of the $\delta N$ expansion,
that is all loop corrections because there is a possibility that higher order loop corrections
give a large contribution in the evaluation of $f_{NL}$ and $\tau_{NL}$. 
We address this question in this paper. 
We will show that the same expression with the original SY inequality is still satisfied approximately as far as the curvature perturbation 
$\delta N$ is expanded by linear scalar fields (namely, no vector fields).

The organization of the present paper is as follows.
In Sec. II, we explain some notations used in this paper.
In Sec. III, we explain about the $\delta N$ formalism briefly.
Then, we show the detailed calculation for the extended of SY inequality up to 1-loop and see  
the same expression with the original SY inequality applies in Sec. IV.  
In Sec V, we generalize the discussion in Sec. IV up to all order in the $\delta N$ expansion.
In Sec VI, we summarize our conclusion.
Detailed calculations in this paper are summarized in the Appendix A, B and C. 

\section{Notations}
We explain some notations used in this paper.

In order to express a tensor components, we use the lower-case alphabet indices such as $a, b, c \dots$, running 1 to $m$.
Then, we express a $n$-rank tensor as
\begin{equation}
		M_{a_1a_2\dots a_n} \equiv M_{a_n} \equiv M_n,
		\label{}
\end{equation}
and the contraction for arbitrary two $n$-rank tensor $M_n$ and $L_n$ is expressed by using a metric ${\cal G}$ as ,
\begin{align}
	 \langle M_{n} , L_{n} \rangle_{ {\cal G}^n } &\equiv	M_{a_n} L_{a_n} \notag \\
	&\equiv	{\cal G}^{a_1 b_1} {\cal G}^{a_2 b_2} \dots {\cal G}^{a_n b_n}  M_{a_1 a_2 \dots  a_n} L_{b_1 b_2 \dots b_n} 
		\label{}
\end{align}
where  $\langle  \cdot , \cdot \rangle_{  {\cal G}^n}$ denotes 
an inner product made by the outer product of metrics, ${\cal G}^n \equiv {\cal G} \otimes {\cal G} \otimes \dots $,
for $n$-rank tensors $M_n$ and $L_n$.

Furthermore, we define  $m(m^n-1)/(m-1)$-dimensional vectors $M$ and $L$ having $n$-rank tensors $M_n$ and $L_n$ as their components,
\begin{align}
	  	L & \equiv \left( L_1, L_2, \dots , L_n \right)   =     \left( L_a , L_{ab}, \dots L_{a_1 a_2 \dots a_n} \right), \notag \\
	    M &\equiv \left( M_1,\ M_2, \dots ,\ M_{n} \right)  =\left( M_a, M_{ab}, M_{a_1 a_2 \dots a_n}\right)  ,
		\label{}
\end{align}
and then we define the inner production for $M$ and $L$ as
\begin{align}
		\langle L, M \rangle_G
	   &\equiv (L_a,L_{ab},L_{abc}, \dots )
        \begin{pmatrix}
			   {\cal G} & 0  & 0 & 0 \\
			   0    & {\cal G} \otimes {\cal G}  & 0   &  0      \\
			   0     & 0     &   {\cal G} \otimes {\cal G} \otimes {\cal G}   &  0  \\
			   0     & 0     &                          0  & \dots 
		\end{pmatrix}
         \begin{pmatrix}
				 M_{a} \\ M_{ab} \\  M_{abc} \\ \dots 
		\end{pmatrix}  
		\notag \\
		& = \sum_{n} \langle L_n, M_n \rangle_{ {\cal G}^n }
		\label{}
\end{align}
where $\langle \cdot  , \cdot \rangle_G$ denotes a inner product defined by the metric $G$ 
which is defined as 
\begin{align}
		G &\equiv {\cal G} \oplus ({\cal G}\otimes{\cal G}) \oplus ({\cal G}\otimes{\cal G}\otimes{\cal G}\oplus ) \oplus \dots \notag \\
          & = \begin{pmatrix}
			   {\cal G} & 0  & 0 \\
			   0         & {\cal G} \otimes {\cal G}     &  0      \\
				0         &         0                           & \dots   
		\end{pmatrix}  .
		\label{}
\end{align}
The metric $G$ is obviously positive-difinite when the metric ${\cal G}$ is positive-difinite.

\section{The $\delta N$ formalism}
In this papar it is assumed that the universe is  dominated by scalar fields in the inflationary period. 
In super horizon scales,
the universe at each points behave as an independent universe, called the {\it separate universe}. 
Furthermore, the primordial curvature perturbation $\zeta$ is described by the fluctuation of the $e$-folding $N \equiv \int_{t_*}^{t} H' dt'$, where $H$ is the Hubble parameter.
Then, the curvature perturbation $\zeta = \delta N$ can be expanded by 
the scalar fields at the initial time which is a few times of horizon crossing time $t_*$ as follows.
\begin{align}
		\zeta = \delta N & = N\left[ \bar{\rho}, \varphi_*^a(\xx) \right] - N\left[ \bar{\rho}, \bar{\varphi}^a_* \right] \notag \\
		& = \frac{\partial N}{\partial \bar{\varphi}^a_*} \delta \varphi^{a}_*(\xx) 
		+ \frac{1}{2} \frac{\partial^2 N}{\partial \bar{\varphi}^a_* \partial \bar{\varphi}^b_*}
		\delta \varphi^a_*(\xx) \delta \varphi_*^b(\xx) + \dots .
		\label{}
\end{align}
where the alphabets such as $a, b, \dots$ denote the number of scalar fields.
Note that in the $\delta N$ formalism,
the $e$-folding must be expressed as a function of the background energy density $\bar{\rho}$ 
and scalar fields evaluated at initial time
\cite{Starobinsky:1982,Starobinsky:1986,Sasaki/Stewart:1996,Lyth/Rodriguez:2005,Valenzuela_Taledo/etal:2011}. 
Here, we express the $\delta N$ expansion as,
\begin{equation}
		\zeta = \delta N = \sum_{n = 1}^{\infty} \frac{1}{n!} N_{a_n} \delta \varphi^{a_n}_*  ,
		\label{}
\end{equation}
where we define as 
\begin{equation}
		\delta \varphi^{a_n}_* \equiv \delta \varphi^{a_1}_* \delta \varphi^{a_2}_* \cdots \delta \varphi^{a_n}_*  ,
		\label{}
\end{equation}
\begin{equation}
		N_{a_n} \equiv N_{a_1 a_2 a_3 \cdots a_n} 
		\equiv \frac{\partial^n N[\bar{\rho},\bar{\varphi}_*^b]}
		{\partial \bar{\varphi}^{a_1}_* \dots \partial \bar{\varphi}^{a_n}_* }  ,
		\label{}
\end{equation}
and $N_{a_n}$ is a complete symmetric tensor.

In the calculation of the $\delta N$ formalism, 
we need the information about initial values of scalar fields.
Therefore, we adopt a following assumption about initial conditions:
\begin{itemize}
		\item Fluctuations in the scalar fields at the horizon crossing $\delta \varphi_{L*}^a$ are scale invariant as well as Gaussian.
   \end{itemize}
The linear power spectrum for the scalar fields at initial time $k \ll (aH)_*$ is assumed to be 
\begin{equation}
		\langle \delta \varphi^a_{L\kk}(t_*) \delta \varphi^b_{L\kk'}(t_*) \rangle
		= (2\pi)^3 \delta_D(\kk + \kk') \frac{2\pi^2}{k^3} {\cal G}^{ab}_{\ast} ,
		\label{}
\end{equation}
where $\delta \varphi_{L\kk}^a(t_*)$ denotes the Fourier transformation of scalar field perturbations.
Since the linear perturbations of scalar fields have Gaussian distribution,
the Non-Gaussianity of the curvature perturbation will
arise from non-linear terms in the $\delta N$ expansion on super horizon scales.
While usually the power spectrum $\mathcal{G}_{\ast}^{ab}$ is assumed to be proportional to  $ \delta^{ab}$, 
we do not adopt this assumption in this paper.
This allow us to treat the case where two fields have a correlation each other at the initial time.
The initial power spectrum $\mathcal{G}^{ab}_{*}$ is a real positive symmetric matrix
and behaves as 
a positive metric in contracting for coefficients of the $\delta N$ formalism.
In a diagrammatical picture of the $\delta N$ formalism,
the non-dimensional initial power spectrum ${\cal G}_*$ plays a role of the coupling constant and {\it n-loop} contributions 
contains the terms proportional to ${\cal G}_*^n$.
Then, the diagrams of the power spectrum, bispectrum and trispectrum are expressed 
by using the linear power spectrum as propagators (see Appendix A and B).

Although the $\delta N$ formalism could be used for vector fields, 
we focus only on scalar fields in this paper,
because the power spectrum for vector fields usually have strong scale-dependence and our proof in the below 
might not apply.
The $\delta N$ formalism and the consistency relation for vector fields 
are studied in \cite{Dimopulos/etal:2009,Beltran/etal:prep}.

\section{Tree level SY inequality}
As in the case of the original SY inequality, which is the most simple case,
we first consider non-Gaissianity parameters $f_{NL}$ and $\tau_{NL}$
arising from the terms of the order $\delta \varphi^2$ in the $\delta N$ expansion.
Namely we consider the following form of the $\delta N$ expansion.
\begin{equation} 
\zeta = N_a \delta \varphi^a_* + \frac{1}{2} N_{ab} \delta \varphi^a_* \delta \varphi^b_* .
\end{equation} 
This corresponds to the tree level calculation in a diagrammatical evaluation of the $\delta N$ formalism.
It has been shown that the following simple forms of the power spectrum, bispectrum and trispectrum 
are obtained in this case, respectively,
\begin{equation}
		{\cal P}_{\zeta} = {\cal P}_*^{ab}N_aN_b \equiv N_aN_a  ,
		\label{}
\end{equation}
\begin{equation}
		 \frac{6}{5} f_{NL} = \frac{N_aN_bN_{ab}}{(N_cN_c)^2}  ,
		\label{}
\end{equation}
\begin{equation}
		\tau_{NL} = \frac{N_aN_{ab}N_{bc}N_c}{(N_dN_d)^3}  ,
		\label{}
\end{equation}
where the coefficients of the $\delta N$ formalism, such as $N_a$, $N_{ab}$, etc., 
are contracted by using the power spectrum of scalar fields ${\cal G}_*^{ab}$.

Here, we use the Cauchy-Schwarz inequality.
When we define the inner product for arbitrary two vectors $v_a$ and $u_a$ as
$\langle v,u \rangle \equiv {\cal G}_*^{ab}v_a u_b$,
then the Cauchy-Schwarz inequality leads to $\langle v,u \rangle^2 \leq \langle v,v \rangle \langle u,u \rangle$.
Choosing $v_a$ and $u_a$ as $v_a \equiv N_a$ and $u_a \equiv N_b N_{ba}$,
we immediately find the original SY inequality.  
\begin{align}
		\left( \frac{6}{5} f_{NL} \right)^2
	&	= \frac{\left \langle N_a, N_{ab}N_b \right \rangle^2}{(N_cN_c)^4}	 \notag \\
	&   \leq \left( \frac{\langle N_a,N_a \rangle \langle   N_{bc}N_c, N_{bd}N_{d} \rangle  }{(N_eN_e)^4} \right) = \tau_{NL}  .
		\label{}
\end{align}

\section{1-loop SY inequality}
In this section, we extend the SY inequality up to 1-loop order. This corresponds to the case where the $\delta N$ expansion is truncated at the order of $\delta\varphi^4$, namely, 
\begin{align} 
\zeta = &  N_a \delta \varphi^a_{\ast} + \frac{1}{2}N_{ab} \delta \varphi^a_{\ast} \delta \varphi^b_{\ast}  \notag \\
& + \frac{1}{3!} N_{abc} \delta \varphi^a_{\ast} \delta \varphi^b_{\ast} \delta \varphi^c_{\ast}
+ \frac{1}{4!} N_{abcd} \delta \varphi^a_{\ast} \delta \varphi^b_{\ast} \delta \varphi^c_{\ast} \delta \varphi^d_{\ast} .
\end{align}
Thus, we shall ignore the contributions in the power spectrum, bispectrum, or trispectrum coming from terms of $O(\delta\varphi^5)$. 
The 4th-order term is necessary when we calculate all of the
1-loop contributions in $f_{\rm NL}$ and $\tau_{\rm NL}$.
Up to this order, not only all of the 1-loop contributions are included, but also 
some of the higher-order loop contributions are included.

Then, the power spectrum is given, up to the 4th order, by
\begin{align}
		 \mathcal{P}_{\zeta}(k)  = &   N_aN_a  + N_{ab}N_{ab} \ln(k L) + N_aN_{abb}\ln(k_{\rm max} L) \notag \\
 & +  \frac{1}{4}N_{acc}N_{abb} \ln^2(k_{\rm max} L)  +  N_{abcc}N_{ab} \ln(k L) \ln(k_{\rm max} L) \dots ,
\label{power}
\end{align}
where the 1st term is the tree contribution, the 2nd and 3rd terms
are the 1-loop contributions, and the 4th and 5th terms are the 2-loop
contributions. 
The $k_{\rm max}$ is the ultra-violet cutoff.
The $L$ is a finite size of a box which is chosen to be much larger than
the region of interest, namely $L$ satisfies the condition $L k \gg 1$ for any interest $k$.
As pointed out in \cite{Lyth:2007}, the $L$ is not large enough to satisfy $\ln(k L) \gg 1$, but it is the order of $1/H_0\sim k_0 =0.002~{\rm Mpc}^{-1}$,
where $H_0$ is the current Hubble parameter and $k_0$ is the usual normalization scale used by the {\rm WMAP} collaboration. 

This result can be simplified by defining the following quantities 
(see Eq.~(25) of \cite{Byrnes/etal:2007}):
\begin{align}
&\tilde{N}_a \equiv N_a + \frac{1}{2}N_{abb} 
 \ln(k_{\rm max} L), \\ 
&\tilde{N}_{ab} \equiv N_{ab} + \frac{1}{2}N_{abcc} 
 \ln(k_{\rm max} L) , \\
 & \tilde{N}_{abc} \equiv N_{abc} + \frac{1}{2} N_{abcdd} \ln (k_{\rm max} L) .
\end{align}
Then Eq.~(\ref{power}) becomes
\begin{equation}
\label{eq:pzeta}
\mathcal{P}_{\zeta}(k) =  \tilde{N}_a \tilde{N}_a  + \tilde{N}_{ab} \tilde{N}_{ab} \ln(k L) + \dots 
\end{equation}
Here, the dots in Eq.~(\ref{eq:pzeta}) include the higher-order terms
such as $N_{abcc}^2$.
This is a nice way of writing the power spectrum etc., as the results
do not include the ultra-violet cutoff, $k_{\rm max}$, explicitly: the
cutoff can be absorbed by redefining the derivatives of $N$.
From now on, we shall remove the tildes from the equations, i.e.,
$\tilde{N}\rightarrow N$. 

As shown in Appendix A, 
the bispectrum and trispectrum with 1-loop corrections
are calculated to be 
\begin{align}
		\frac{6}{5}f_{NL}(k_{b_1},k_{b_2}) = {\cal P}^{-1}_{\zeta}(k_{b_1})  {\cal P}^{-1}_{\zeta}(k_{b_2}) 
		\Big[N_aN_{ab}N_b + N_aN_{abc}N_{bc}\left( \ln(k_{b_1} L) + \ln(k_{b_2} L) \right) 
		+ N_{ab}N_{bc}N_{ca} \ln\left( k_{b_1} L \right)  \Big]  ,
		\label{fNL}
\end{align}
\begin{align}
		\tau_{NL}(k_{t_1},k_{t_2}) = &{\cal P}_{\zeta}^{-1}(k_{t_1})  {\cal P}_{\zeta}^{-2}(k_{t_2}) 
		\Big[ N_aN_{ab}N_{bc}N_c + 2 N_{ab}N_{abc}N_{cd}N_d \ln(k_{t_{2}} L)   \notag \\
		& \hspace{3cm} + \Big( N_{ab}N_{bc}N_{cd}N_{da} + N_aN_{abc}N_{bcd}N_d +  2 N_aN_{abc}N_{bd}N_{dc} \Big)\ln (k_{t_1} L )  \Big] ,
		\label{tNL}
\end{align}
where we have used  {\it pole approximation} to evaluate the main contribution of the loop integral(see Appendix A for more detail). The scale dependence is satisfied by the following conditions;
$k_{b_1} \ll k_{b_2}$ and $k_{t_1} \ll k_{t_2}$.
These expressions have been again simplified by using the redefinition of the
derivatives of $N$ and ignoring the higher-order terms.
Here, we can set that $k_{b_1} = k_{t_1} \equiv k_L$ and $k_{b_2} = k_{t_2} \equiv k_S$ without loss of generality,
because the squeezed limit is satisfied for the bispectrum also in this setting.
Therefore, we might focus only on the relation between $f_{NL}(k_{L},k_{S})$ and $\tau_{NL}(k_{L},k_{S})$.

Now we are ready to prove the SY inequality. First we use the following inequality between arbitrary tensors such as $M_a$, $M_{ab}$ and $L_a$, $L_{ab}$,
which is an application of the Cauchy-Schwarz inequality,
\begin{align}
		&\left( M_aL_a  + M_{ab}L_{ab} \right)^2 \le\left(M_aM_a + M_{ab} M_{ab} \right) \left( L_aL_a + L_{ab}L_{ab} \right),
		\label{}
\end{align}
where a positive-definite metric is used in contracting the indices of tensors.

Choosing $M_a$, $M_{ab}$ and $L_a$, $L_{ab}$ as follows.
\begin{align}
		& M_a \equiv N_a  ,\notag \\ 
		& M_{ab} \equiv N_{ab} \left[ \ln (k_L L) \right]^{1/2}, \notag \\ 
		& L_{a} \equiv N_{ab}N_{b} + N_{abc}N_{bc} \ln (k_S L) , \notag \\
		& L_{ab} \equiv \left(N_{abc}N_c + N_{ac}N_{cb}\right)\left[ \ln (k_L L) \right]^{1/2},
		\label{}
\end{align}
Then it is easy to show the following inequality. 
\begin{align}
		&\left( \frac{6}{5}f_{NL}\left( k_L,k_S \right) \right)^2 \notag \\
		&\leq \tau_{NL}(k_L,k_S) 
		+ \frac{N_{ab}N_{abc}N_{cde}N_{de} \left[ \ln (k_S L) \right]^2}{{\cal P}_{\zeta}(k_{L}) {\cal P}_{\zeta}^2(k_{S})} ,
		\label{0th}
\end{align}
where the last term in the right-hand-side (RHS) of Eq.~(\ref{0th}) is the ``2~loop'' term.
This result shows that, when we allow ourselves for completely general
models in which this particular 2-loop term can become important, the
SY inequality, 
$\tau_{\rm NL}\ge  \left( \frac{6}{5}f_{\rm NL}
\right)^2$, may be violated badly. 
The reason of this violation  is 
that we trancated the $\delta N$ expansion at a finite order.

Still, from a model-building point of view, it is reasonable to assume
that the 2-loop terms are sub-dominant compared to the tree or 1-loop
terms; otherwise, we would have to require fine-tunings between the
derivatives of $N$.
Let us then study the consequence of ignoring this particular 2-loop term.
We shall impose the following condition:
\begin{align}
		\frac{N_{ab}N_{abc}N_{cde}N_{de} [\ln(k_{S}L)]^2 }{N_{b}N_{ba}N_{ac}N_c} 
		\ll  \Bigg| \frac{N_dN_{da}N_{abc}N_{bc} \ln(k_{S}L)}{N_bN_{ba}N_{ac}N_c}\Bigg|.
\label{con}
\end{align}
This condition is 
(1-loop)$\gg$(2-loop) for the second term in the RHS of Eq.~(\ref{tNL}) and 
the last term in the RHS of Eq.~(\ref{0th}).
Interestingly, from the Cauchy-Schwarz inequality for $v_a \equiv N_{ab}N_b$ and
$u_a \equiv N_{abc}N_{bc}$, we find
\begin{align}
		\Bigg( \frac{N_dN_{da}N_{abc}N_{bc} \ln(k_{S} L)}{N_bN_{ba}N_{ac}N_c} \Bigg)^2 
		\leq \frac{N_{ab}N_{abc}N_{cde}N_{de} [\ln(k_{S}L )]^2 }{N_{b}N_{ba}N_{ac}N_c} ,
 \label{rel}
\end{align}
and from Eq.~(\ref{con}) we find 
\begin{equation}
		\Bigg| \frac{N_dN_{da}N_{abc}N_{bc} \ln(k_{S}L)}{N_bN_{ba}N_{ac}N_c} \Bigg|^2 \ll 
		\Bigg| \frac{N_dN_{da}N_{abc}N_{bc} \ln(k_{S}L)}{N_bN_{ba}N_{ac}N_c} \Bigg| .
		\label{}
\end{equation}
Finally, we obtain the following restriction on a particular form of 1-loop contributions: 
\begin{equation}
		\Bigg| \frac{ N_dN_{da}N_{abc}N_{bc} \ln(k_{S}L) }{N_bN_{ba}N_{ac}N_c} \Bigg| \ll 1.
\label{aho}
\end{equation}
As a result, if we ignore the last term in the RHS of Eq.~(\ref{0th}),
we must also ignore the second term in the RHS of Eq.~(\ref{tNL}). This is a peculiar feature of these terms,
whose physical meaning is not clear. 

In any case, provided that the following additional condition is met:
\begin{itemize}
\item  The 2-loop contributions are sub-dominant compared to the
       1-loop contributions (or at least the particular
      2-loop term in 
	  the RHS of Eq.~(\ref{0th}) is small compared to the others),
\end{itemize}
we finally arrive at the extended version of the SY inequality up to 1-loop:
\begin{equation}
		\tau_{\rm NL}(k_{L},k_{S})
		\geq \left( \frac{6}{5}f_{\rm NL} (k_{L},k_{S})\right)^2,
  \label{}
\end{equation}
which is valid as long as the 2-loop contributions are small.
As pointed out in \cite{C.T.Byrnes/etal:prep},
we do not have the equality in the cases of single field models (single degree of freedom models) any more.

It is thus proved that
the same expression with the original SY inequality is satisfied even with 1-loop corrections.
This result is a stronger than one in our previous work \cite{N.S.Sugiyama/Komatsu/etal:2011},
and the previous conclusion that the 1-loop corrections weaken the original inequality should be modified.

\section{Generalization up to any loop SY inequality}
Now we prove the most general case where all order of the loop corrections are taken into account.   The detail of the calculation for the power spectrum, bispectrum and trispectrum with any loop corrections will be given in the Appendix B.

The power spectrum including all loop corrections is
\begin{align}
		{\cal P}_{\zeta}(k)   & = \sum_{n=1}^{\infty} \left( \frac{n}{n!} \right) 
		 N_{a_n}  N_{a_n} (\ln kL)^{n-1} \notag \\ 
		 & = \sum_{n=1}^{\infty} \left( \frac{n}{n!} \right) 
		 \langle N_{n} , N_{n} \rangle_{ {\cal G}^n}(\ln kL)^{n-1}  \notag \\
		 & = \sum_n \langle P_n(k), P_n(k) \rangle_{ {\cal G}^n  } = \langle P(k),P(k) \rangle_G ,
		\label{general_power}
\end{align}
where we defined $n$-rank tensor $P_n$ depending on $k$ as
\begin{equation}
		P_n(k) \equiv\sqrt{\frac{n}{n!}} (\ln kL)^{\frac{n-1}{2}}   N_n .
		\label{}
\end{equation}
The $f_{NL}$ and $\tau_{NL}$ with all loop contributions are given by
\begin{align}
	    \frac{6}{5}f_{NL}(k_{b_1},k_{b_2})  = & 
		\frac{1}{ {\cal P}_{\zeta}(k_{b_1}) {\cal P}_{\zeta}(k_{b_2})} 
		 \sum_{m_1 = 1}^{\infty} \sum_{m_2=0}^{\infty} \sum_{m_3=1}^{\infty} 
		 		[\ln k_{b_1} L]^{m_2+m_3-1} [\ln k_{b_2} L]^{m_1-1}\notag \\
			&\hspace{3cm}\times \frac{m_1m_3}{ m_1! m_2! m_3!} 
          N_{a_{m_1} b_{m_2}} N_{b_{m_2} c_{m_3}} N_{c_{m_3} a_{m_1}} ,
		\label{}
\end{align}
\begin{align}
		\tau_{NL}(k_{t_1},k_{t_2})
		=& \frac{1}{ {\cal P}_{\zeta}(k_{t_1})  {\cal P}^2_{\zeta}(k_{t_2}) }
         \sum_{m_1 = 1}^{\infty} \sum_{m_2=1}^{\infty}  \sum_{m_3=1}^{\infty} 
		 \sum_{m_4=0}^{\infty}  \sum_{l_1=0}^{\infty}  \sum_{l_2=0}^{\infty}
		[\ln k_{t_1}L]^{m_2 + m_4+ l_1 + l_2-1} [\ln k_{t_2} L]^{m_1 + m_3-2}
		    \notag \\
			&\hspace{2.5cm} \times\frac{m_1m_2m_3}{m_1!m_2!m_3!m_4!l_1!l_2!} 
		 N_{a_{m_1}f_{l_2}d_{m_4}}N_{a_{m_1}e_{l_1}b_{m_2}}N_{b_{m_2}f_{l_2}c_{m_3}}N_{c_{m_3}e_{l_1} d_{m_4}} ,
		 		\label{}
\end{align}
where the squeezed limit $k_{b_1} \equiv k_1 \ll k_2 \simeq k_3 \equiv k_{b_2}$ and
the collinear limit $k_{t_1} \equiv |\kk_1 + \kk_2| \ll k_1 \simeq k_2 \simeq k_3 \simeq k_4 \equiv k_{t_2}$
are taken, respectively.
Even in this case, we can assume $k_{b_1} = k_{t_1} \equiv k_L$ and $k_{b_2} = k_{t_2} \equiv k_S$, where $k_L \ll k_S$, without loss of generality.

Now, we are ready to derive the consistency relation between $f_{NL}(k_L,k_S)$ and $\tau_{NL}(k_L,k_S)$. 

First, we express $f_{NL}$ in terms of $\langle \cdot ,\cdot  \rangle_G$.
For this purpose,
we express $f_{NL}$ as
\begin{align}
		 \frac{6}{5}f_{NL}(k_{L},k_{S})  & = \frac{1}{ {\cal P}_{\zeta}(k_{L}) {\cal P}_{\zeta}(k_{S})} 
		\sum_{\alpha=1}^{\infty} \sum_{m_1 = 1}^{\infty}  \sum_{m_2=0}^{\alpha-1} 
		 [\ln{k_{L} L}]^{\alpha-1} [\ln{k_{S} L}]^{m_1-1} \notag \\
		 & \hspace{1cm} \times \frac{m_1(\alpha-m_2)}{ m_1! (\alpha-m_2)! m_2!}
		  N_{a_{m_1} b_{m_2}} N_{b_{m_2} c_{\alpha-m_2}} N_{c_{\alpha-m_2} a_{m_1}}
		 \notag \\
		& =
         \frac{1}{ {\cal P}_{\zeta}(k_{L}) {\cal P}_{\zeta}(k_{S})} 
		\sum_{\alpha=1}^{\infty} \sum_{m_1 = 1}^{\infty}  \sum_{m_2=0}^{\alpha} 
		 [\ln{k_{L} L}]^{\alpha-1} [\ln{k_{S} L}]^{m_1-1} \notag \\
		 & \hspace{1cm} \times  \frac{m_1(\alpha-m_2)}{ m_1! (\alpha-m_2)! m_2!}
		  N_{a_{m_1} b_{m_2}} N_{b_{m_2} c_{\alpha-m_2}} N_{c_{\alpha-m_2} a_{m_1}} \notag \\
		  & =  \frac{1}{ {\cal P}_{\zeta}(k_{L}) {\cal P}_{\zeta}(k_{S})} 
		  \sum_{\alpha=1}^{\infty} \sum_{m_1 = 1}^{\infty}  \sum_{\tilde{m}_2=0}^{\alpha} 
		 [\ln{k_{L} L}]^{\alpha-1} [\ln{k_{S} L}]^{m_1-1} \notag \\
		 & \hspace{1cm} \times  \frac{m_1\tilde{m}_2}{ m_1! (\alpha-\tilde{m}_2)! \tilde{m}_2!}
		 N_{a_{m_1} b_{\alpha-\tilde{m}_2}} N_{b_{\alpha-\tilde{m}_2} c_{\tilde{m}_2}} N_{c_{\tilde{m}_2} a_{m_1}} ,
		\label{ff}
\end{align}
where $\alpha$ is defined as $\alpha \equiv m_3 + m_2$.
In the second equality,
the upper limit of $m_2$ is replaced from $\alpha-1$ into $\alpha$,
because for $m_2 = \alpha$ the right-hand-side (RHS) of Eq.~(\ref{ff}) becomes zero.
We have used $\tilde{m}_2 = \alpha-m_2$ in the third equality.
Then, we find 
\begin{align}
		\frac{6}{5}f_{NL}(k_L,k_S) & =  \frac{1}{ {\cal P}_{\zeta}(k_{L}) {\cal P}_{\zeta}(k_{S})} 
		\sum_{\alpha=1}^{\infty} \sum_{m_1 = 1}^{\infty}  \sum_{m_2=0}^{\alpha} 
		 [\ln{k_{L} L}]^{\alpha-1} [\ln{k_{S} L}]^{m_1-1} \notag \\
		 & \hspace{1cm} \times \frac{1}{2}\Bigg[ \frac{m_1(\alpha-m_2)}{ m_1! (\alpha-m_2)! m_2!}
		  + \frac{m_1m_2}{ m_1! (\alpha-m_2)! m_2!} \Bigg]
		  N_{a_{m_1} b_{m_2}} N_{b_{m_2} c_{\alpha-m_2}} N_{c_{\alpha-m_2} a_{m_1}} \notag \\ 
		  & = \frac{1}{ {\cal P}_{\zeta}(k_{L}) {\cal P}_{\zeta}(k_{S})} 
		\sum_{\alpha=1}^{\infty} \sum_{m_1 = 1}^{\infty}  \sum_{m_2=0}^{\alpha} 
		 [\ln{k_{L} L}]^{\alpha-1} [\ln{k_{S} L}]^{m_1-1} \notag \\
		 & \hspace{1cm} \times  \frac{1}{2} \frac{m_1\alpha}{ m_1! (\alpha-m_2)! m_2!}
		  N_{a_{m_1} b_{m_2}} N_{b_{m_2} c_{\alpha-m_2}} N_{c_{\alpha-m_2} a_{m_1}}.
		\label{}
\end{align} 

Choosing $M_{\alpha}$ as
\begin{equation}
		M_{\alpha}(k_{L},k_{S})
		 \equiv \sum_{m_1 = 1}^{\infty}  \sum_{m_2=0}^{\alpha} 
		 \frac{1}{2}\frac{m_1 \sqrt{\alpha} \sqrt{\alpha !}}{ m_1! (\alpha-m_2)! m_2!}(\ln{k_{L} L})^{\frac{\alpha-1}{2}}
         (\ln{k_{S} L})^{m_1-1}
         N_{ b_{m_2}a_{m_1}}  N_{a_{m_1} c_{\alpha-m_2}}  ,
		\label{}
\end{equation}
and we find
\begin{align}
	\left( \frac{6}{5}f_{NL}(k_{L},k_{S}) \right)	
	& =  \frac{1}{ {\cal P}_{\zeta}(k_{L}) {\cal P}_{\zeta}(k_{S})} 
	\sum_{\alpha=1}^{\infty} \langle P_{\alpha}(k_{L}) , M_{\alpha}(k_{L},k_{S}) \rangle_{ {\cal G}^{\alpha}} \notag \\
	& = \frac{1}{ {\cal P}_{\zeta}(k_{L}) {\cal P}_{\zeta}(k_{S})} 
	\left\langle P(k_{L}), M(k_{L},k_{S}) \right \rangle_G ,
		\label{}
\end{align}
where $M_{\alpha}$ is a symmetrized completely
because it is contracted by the complete symmetric tensor $P_{\alpha}$.

Next, we can use the Cauchy-Shwarz inequality for the inner product $\langle \cdot , \cdot \rangle_G$
because the inner product is positive-difinite,
and we find
\begin{align}
		\left( \frac{6}{5}f_{NL}(k_{L},k_{S}) \right)^2
		& = \frac{1}{ {\cal P}^2_{\zeta}(k_{L}) {\cal P}^2_{\zeta}(k_{S})} 
        \left\langle P(k_{L}), M(k_{L},k_{S}) \right \rangle_G^2 \notag \\
        & \leq \frac{1}{ {\cal P}^2_{\zeta}(k_{L}) {\cal P}^2_{\zeta}(k_{S})} 
		\left\langle P(k_{L}), P(k_{L}) \right \rangle_G
		\left \langle M(k_{L},k_{S}),M(k_{L},k_{S}) \right\rangle_G \notag \\
		& = \frac{1}{ {\cal P}_{\zeta}(k_{L}) {\cal P}^2_{\zeta}(k_{S})} 
		\left \langle M(k_{L},k_{S}),M(k_{L},k_{S}) \right\rangle_G   \notag \\
		& = \tau_{NL}(k_L,k_S)
		\label{C-S_inequality}
\end{align}
where the last equality will be shown in Appendix C.

We finally arrive at main result:
\begin{equation}
		\tau_{NL}(k_{L},k_{S}) \geq \left(\frac{6}{5} f_{NL}(k_{L},k_{S}) \right)^2 .
		\label{main}
\end{equation}
This inequality has the same expression with the original SY inequality with only one difference.
Namely, the equality is no longer satisfied even in the case of single field models.

Note that this inequality contains some errors.
First, we use the pole approximation in the evaluation of the loop integrals. Exactly speaking, the power spectrum, bispectrum and trispectrum calculated by using the pole approximation is different from the truly evaluated quantities.
In fact, the investigation of the validity of the pole approximation is an interesting problem.
In \cite{Kumar/Leblond/etal:2010}, 
it is shown that the pole approximation is indeed valid for bispectrum with 1-loop corrections in the leading order.
Furthermore, there is also a problem
about the infrared divergences which are eliminated by introducing the size $L$ of the observed patch of the late universe \cite{Gerstenlauer:2011ti,Seery:2010kh}.
Second, theoretically predicted bispectrum and trispectrum with all loop corrections
do not coincide with the local form  bispectrum and trispectrum used in the observation. 
Therefore, we take the squeezed limit $k_L \equiv k_1 \ll k_2 \simeq k_3 \equiv k_S$ for  $f_{NL}$ and the collinear limit 
$k_L \equiv |\kk_1 + \kk_2| \ll k_1 \simeq k_2 \simeq k_3 \simeq k_4 \equiv k_S$ for $\tau_{NL}$ to make the calculated bispectrum and trispectrum into the local forms approximately. While the obtained limits of the bispectrum and trispectrum are approximately local form and have the maximum values at these limits as in the case of the local form bispectrum and trispectrum, we cannot completely remove some errors.

Nevertheless,
the inequality Eq.~(\ref{main}) would be still useful for focusing on the observation of a complete violation of the inequality, 
because the errors are considered not to be large enough to change our inequality, i.e.,
\begin{equation}
\tau_{NL} \ll \left( \frac{6}{5}f_{NL} \right)^2 .
		\label{}
\end{equation}

\section{Conclusion}
We calculated the power spectrum, birpsctrum and trispectrum including the contributions of all higher order loop corrections for general multi-field 
models of inflationary scenario using $\delta N$ formalism. 
Then, we derived the following consistency relation between non-linear parameters under the assumption that the pole approximation dominates the higher order loop integral.
\begin{equation}
\tau_{NL}(k_L,k_S) \geq \left( \frac{6}{5}f_{NL}(k_L,k_S) \right)^2.
\end{equation} 
This is exactly the same expression with the original SY inequality.
However, there is one difference.
While the equality is satisfied for single field models in the original SY inequality,
we do not have the equality any more in the extended SY inequality.

There are some errors for the extended SY inequality.
The uncertainty causes the difference between theoretically calculated power spectrum, bispectrum, and trispectrum
and the corresponding spectrums in the observation.
We have to use the pole approximation to treat the loop integral analytically,
and further take the squeezed limit and collinear limit to 
make the calculated bispectrum and trispectrum by using the pole approximation 
into the observed non-linear parameters $f_{NL}$ and $\tau_{NL}$.
Although the dominant contributions to the non-linear parameters would satisfy the SY inequality,
these errors might slightly violate the SY inequality.

Therefore, the most interesting case is the observation 
of a complete violation of the inequality, that is 
$\tau_{NL}\ll \left( \frac{6}{5}f_{NL} \right)^2$.
Then the observation implies that a large class of inflation models
are rejected for the mechanism to generate the observed fluctuations,
{\it provided} that 
(1) scalar fields generate the primordial curvature perturbation;
(2) the $\delta N$ formalism can be applied;
(3) fluctuations in the scalar fields at the horizon crossing are scale invariant as well as Gaussian.

\acknowledgments
We would like to thank T. Futamase for useful discussion and comments.
This work is supported in part by the GCOE Program ``Weaving Science Web beyond
Particle-matter Hierarchy'' at Tohoku University and
by a Grant-in-Aid for Scientific Research from JSPS 
(Nos. 18072001, 20540245 for TF) as well as by Core-to-Core Program
``International Research Network for Dark Energy.''
TF thanks to Luc Branchet and Institute of Astronomical Observatory, Paris for warm hospitality during his stay in the last stage 
of the present work.

\appendix

\section{Loop integrals: 1-loop}
In this appendix, 
we show the detail calculations of the 1-loop integrals.
Here, the linear power spectrum of scalar fields, which is behave as a propagator, is given by
\begin{equation}
		 \langle \delta \varphi^a_{\kk}(t_{\ast}) \delta\varphi^b_{\kk^{\prime}}(t_{\ast}) \rangle 
		 = (2\pi)^3 \delta^{(3)}(\kk + \kk^{\prime})\frac{2\pi^2}{k^3}\mathcal{G}_{\ast}^{ab}.
		\label{}
\end{equation}
where ${\cal G}_*^{ab}$ is the non-dimensional power spectrum and
behaves as a positive metric in contracting for coefficients of the $\delta N$ formalism.
The diagrams for non-linear parameters $f_{NL}$ and $\tau_{NL}$ is shown in Fig.~(\ref{fig:fNL-1loop}) and Fig.~(\ref{fig:tau-1loop}),
respectively.

\subsection{power spectrum}
The power spectrum including a loop contribution is described as
\begin{equation}
		{\cal P}_{\zeta}(k) = N_aN_a + N_{ab}N_{ab}\frac{k^3}{8\pi} \int_{1/L} d^3p \frac{1}{|\kk-\pp|^3} \frac{1}{p^3} ,
		\label{}
\end{equation}
where the subscript $L$ on the integral denotes
that the integrand is set equal to zero at $p < L^{-1}$.
This setting is necessary in order to cut out a sphere around each of the singularities of the integrand.
These singularities yield from the flatness of the linear power spectrum of scalar fields at initial time ${\cal P}_*$,
diverging logarithmically at the singularities.
This loop integral is analytically calculable,
\begin{align}
		&\int_{1/L} d^3p \frac{1}{|\kk-\pp|^3} \frac{1}{p^3} \notag \\
		& = 4\pi\Bigg(\int_{1/L}^{k-1/L}\frac{dp}{kp}\frac{1}{k^2-p^2} + \int_{k + 1/L }^{\infty} \frac{dp}{p^2} \frac{1}{p^2-k^2}\Bigg)\notag \\
		& \sim \frac{4\pi}{k^3}(2\ln(kL) -1)   \sim \frac{8\pi}{k^3} \ln(k L),
		\label{power_int}
\end{align}
where we have used $k L \gg 1$.
The UR divergence do not appear in this calculation.
If we assume as $\ln(k L ) \sim {\cal O}(1)$ with $L \sim k_0$,
the last term in the second line in the RHS of Eq.(\ref{power_int})
might not be neglected strictly,
but we contain the term  into the $\ln(k L)$ as the uncertainty of the order unity.
We can regard this result as 
the factor $\ln (kL)$ yield from each singularities, such as $\pp = 0$ and $\pp =\kk$.

Finally, we can find 
\begin{equation}
		{\cal P}_{\zeta} (k)= N_aN_a + N_{ab}N_{ab} \ln(k L).
		\label{}
\end{equation}

\subsection{bispectrum}
\begin{figure}[t]
		\begin{center}
				\scalebox{1.3}{\psfig{figure=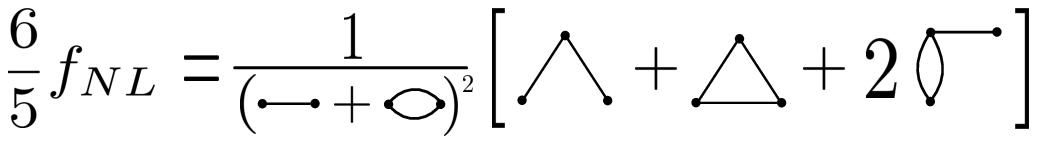}}
		\end{center}
		\caption{Diagrams of the non-linear parameter $f_{NL}$ with 1-loop corrections}
		\label{fig:fNL-1loop}
\end{figure}
The bispectrum with 1-loop integrals is
\begin{align}
		B_{\zeta}(k_1,k_2,k_3)   = &  N_aN_{ab}N_b (2\pi^2)^2
		\Bigg[\frac{1}{k_1^3 k_2^3} + \frac{1}{k_1^3k_3^3} + \frac{1}{k_2^3k_3^3} \Bigg] \notag \\
		& +  N_{ab}N_{bc}N_{ca} (2\pi^2)^3 
		\int_{1/L} \frac{d^3p}{(2\pi)^3}\frac{1}{p^3}\frac{1}{|\kk_1 + \pp|^3} \frac{1}{|\kk_3-\pp|^3}  \notag \\
		& + \frac{N_aN_{abc}N_{bc}}{2}(2\pi^2)^3
		\Bigg[\frac{1}{k_2^3}\int_{1/L}\frac{d^3p}{(2\pi)^3} \frac{1}{|\kk_1-\pp|^3} \frac{1}{p^3} + \mbox{5 perm.} \Bigg].
		\label{bi_int}
\end{align}
The second term in the RHS of Eq.(\ref{bi_int}) is not calculable analytically,
but we assume that the main contributions in the integral yield from the singularities of the integrand as like the case of power spectrum,
called the {\it pole approximation}.
Therefore, this integral is calculated as follows,
\begin{align}
		\int_{1/L} \frac{d^3p}{(2\pi)^3}\frac{1}{p^3}\frac{1}{|\kk_1 + \pp|^3} \frac{1}{|\kk_3-\pp|^3}  
		 = & \int_{1/L} \frac{d^3p}{(2\pi)^3}\frac{1}{p^3}\frac{1}{|\kk_1 + \pp|^3} \frac{1}{|\kk_3-\pp|^3} \Bigg|_{\pp = 0}  \notag \\
		& + \int_{1/L} \frac{d^3q}{(2\pi)^3}\frac{1}{|\qq-\kk_1|^3}\frac{1}{q^3} \frac{1}{|\kk_2 + \qq|^3} 
		\Bigg|_{\qq\equiv \kk_1 + \pp = 0} \notag \\
		& + \int_{1/L} \frac{d^3q}{(2\pi)^3}\frac{1}{|\kk_3-\qq|^3}\frac{1}{|\kk_2 + \qq|^3} \frac{1}{q^3} 
		\Bigg|_{\qq\equiv \kk_3-\pp = 0} \notag \\
		 = & \frac{1}{2\pi^2}\Bigg[\frac{1}{k_1^3} \frac{1}{k_3^3}  \int_{1/L}^{\rm Min\{k_1,k_3\}} \frac{dp}{p} 
		   + \mbox{ 2 perm.} \Bigg]\notag \\
		 = & \frac{1}{2\pi^2}
		\Bigg[ \frac{\ln(\rm Min\{k_1,k_3\} L) }{k_1^3k_3^3} + \mbox{2 perm.}\Bigg] ,
		\label{}
\end{align}
where the upper limit of integral is determined as the next singular point.

On the other hand,
the third term in the RHS of Eq.(\ref{bi_int}) is calculable analytically as like the power spectrum,
and we find
\begin{align}
		&\frac{1}{k_2^3}\int_{1/L} \frac{d^3p}{(2\pi)^3} \frac{1}{|\kk_1-\pp|^3} \frac{1}{p^3} + {\rm 5 perm.}  \notag \\
		&=\frac{2}{2\pi^2}\Bigg[\frac{\ln(k_1 L) + \ln(k_2 L)}{k_1^3k_2^3} + {\rm 2 perm.} \Bigg] .
		\label{}
\end{align}

Finally, we can find the bispectrum contained the 1-loop contributions as follows,
\begin{align}
		 B_{\zeta}(k_1,k_2,k_3) = &
		 \frac{N_aN_{ab}N_b + N_aN_{abc}N_{bc}\left( \ln(k_1 L) + \ln(k_2 L) \right) 
		+ N_{ab}N_{bc}N_{ca} \ln\left( \rm Min \{k_1,k_2\} L \right) }
		{\left(N_aN_a + N_{ab}N_{ab} \ln(k_1 L)\right) (N_aN_a + N_{ab}N_{ab} \ln(k_2 L))} 
		 P_{\zeta}(k_1) P_{\zeta}(k_2)  \notag \\
		 & + \hspace{9cm}  \mbox{ 2 perm.} .
		\label{}
\end{align}
This form does not coincide with the local form bispectrum,
but this form has a maximum value in squeezed limit,
hence we can correspond the 1-loop bispectrum
to the local form $f_{NL}$ by imposing the squeezed limit, $k_{b_1} \equiv k_1 \ll k_2 \sim k_3 \equiv k_{b_2}$. 
Then we find
\begin{align}
		\frac{6}{5}f_{NL}(k_{b_1},k_{b_2}) = & {\cal P}^{-1}_{\zeta}(k_{b_1}) {\cal P}^{-1}_{\zeta}(k_{b_2}) 
		  \Big[N_aN_{ab}N_b + N_{ab}N_{bc}N_{ca} \ln\left( k_{b_1} L \right)  
		  + N_aN_{abc}N_{bc}\left( \ln(k_{b_1} L) + \ln(k_{b_2} L) \right) \Big] .
				\label{}
\end{align}
The diagrams for this $f_{NL}$ are shown in Fig.~(\ref{fig:fNL-1loop}).
\subsection{trispectrum}
\begin{figure}[t]
		\begin{center}
				\resizebox{150mm}{13mm}{\psfig{figure=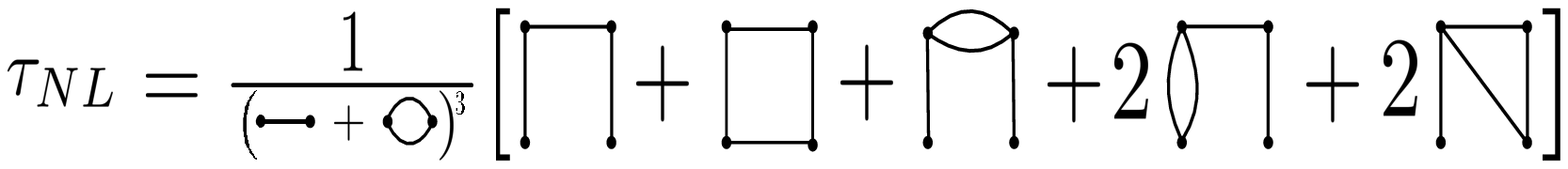}}
		\end{center}
		\caption{Diagrams of non-linear parameter $\tau_{NL}$ with 1-loop corrections.}
		\label{fig:tau-1loop}
\end{figure}
The trispectrum including 1-loop integrals is
\begin{align}
	T_{\zeta}(k_1,k_2,k_3,k_4)  
	= & N_{a}N_{ab}N_{bc}N_c(2\pi^2)^3 \Bigg[ \frac{1}{k_1^3|\kk_1 + \kk_2|^3 k_4^3 } + \mbox{ 12 perm.} \Bigg] \notag \\
	& +   N_{ab}N_{bc}N_{cd}N_{da} (2\pi^2)^4 
	\int_{1/L} \frac{d^3p}{(2\pi)^3} \Bigg(\frac{1}{p^3}\frac{1}{|\kk_1 + \pp|^3} \frac{1}{|\kk_1 + \kk_2 + \pp|^3} \frac{1}{|\kk_4-\pp|^3} 
	 + \mbox{ 2 perm.} \Bigg)\notag \\
	 & + \frac{N_{ab}N_{abc}N_{cd}N_d}{2} (2\pi^2)^4 
	 \Bigg[ \Bigg( \frac{1}{|\kk_1 + \kk_2|^3} \frac{1}{k_4^3} \int_{1/L} \frac{d^3p}{(2\pi)^3}\frac{1}{|\kk_1-\pp|^3}  \frac{1}{p^3} \Bigg)
		 + \mbox{ 23 perm.} \Bigg] \notag \\
		 & + \frac{N_aN_{abc}N_{bcd}N_d}{2} (2\pi^2)^4 
		\Bigg[ \Bigg( \frac{1}{k_1^3}\frac{1}{k_4^3}  \int_{1/L}\frac{d^3p}{(2\pi)^3} \frac{1}{|\kk_1+\kk_2-\pp|^3}  \frac{1}{p^3} \Bigg)
		 + \mbox{ 11 perm.} \Bigg] \notag \\
		 &  + \frac{N_aN_{abc}N_{bd}N_{dc}}{2} (2\pi^2)^4 
		\Bigg[\frac{1}{k_1^3}\int_{1/L} \frac{d^3p}{(2\pi)^3} \frac{1}{p^3}\frac{1}{|\kk_1 + \kk_2+\pp|^3} \frac{1}{|\kk_4-\pp|^3} 
		+ \mbox{ 23 perm.} \Bigg] .
		\label{tau_int}
\end{align}
The second term in the RHS of Eq.(\ref{tau_int}) is not calculable analytically as like the case of bispectrum,
but this integral is evaluated by using the pole approximation as follows,
\begin{align}
		&\int_{1/L}\frac{d^3p}{(2\pi)^3}
		\Bigg(\frac{1}{p^3}\frac{1}{|\kk_1 + \pp|^3} \frac{1}{|\kk_1 + \kk_2 + \pp|^3} \frac{1}{|\kk_4-\pp|^3} 
		+ \mbox{ 2 perm.} \Bigg) \notag \\
		&\hspace{1cm} = \frac{1}{2\pi^2}
		\Bigg[ \frac{\ln \left( \rm Min\{k_1,k_4,|\kk_1 + \kk_2| \} \right)}{k_1^3|\kk_1 + \kk_2|^3 k_4^3} + \mbox{ 11 perm.} \Bigg] .
		\label{}
\end{align}

The third and forth terms are calculable analytically,
and we find respectively,
\begin{align}
		   \Bigg( \frac{1}{|\kk_1 + \kk_2|^3} \frac{1}{k_4^3} \int_{1/L}\frac{d^3p}{(2\pi)^3}
		 \frac{1}{|\kk_1-\pp|^3}  \frac{1}{p^3} \Bigg) + \mbox{ 23 perm.}  
		 = \frac{2}{2\pi^2}\Bigg[ \frac{\ln(k_1 L) + \ln (k_4 L)}{k_1^3 |\kk_1 + \kk_2|^3 k_4^3} + \mbox{11 perm.} \Bigg] ,
		\label{}
\end{align}

\begin{align}
	 \Bigg( \frac{1}{k_1^3}\frac{1}{k_4^3}  \int_{1/L}\frac{d^3p}{(2\pi)^3}
 \frac{1}{|\kk_1+\kk_2-\pp|^3}  \frac{1}{p^3} \Bigg) + \mbox{ 11 perm.}
 = \frac{2}{2\pi^2} \Bigg[\frac{\ln(|\kk_1 + \kk_2| L)}{k_1^3 |\kk_1 + \kk_2|^3 k_4^3} + {\rm 11 perm.} \Bigg] .
		\label{}
\end{align}

The fifth term is calculated by using the pole approximation,
but we need a notice.
In the last term of the RHS in (\ref{tau_int}),
if we set the pole around $\kk_4$,
we do not get the trispectrum corresponding to $\tau_{NL}$.
When we extract only terms corresponding to $\tau_{NL}$,
we find 
\begin{align}
		&\frac{1}{k_1^3}\int_{1/L} \frac{d^3p}{(2\pi)^3} \frac{1}{p^3}\frac{1}{|\kk_1 + \kk_2+\pp|^3} \frac{1}{|\kk_4-\pp|^3} 
		+ \mbox{ 23 perm.}  \notag \\
		& =  \frac{2}{2\pi^2} \Bigg[ \frac{\ln (\rm Min \{k_1,|\kk_1+\kk_2| \}) + \ln(\rm Min \{|\kk_1+\kk_2|,k_4 \})}
		{k_1^3|\kk_1 + \kk_2|^3 k_4^3} + {\rm 11 perm.} \Bigg] .
		\label{}
\end{align}

Finally, we find the trispectrum including 1-loop contributions as follows,
\begin{align}
		T_{\zeta}(k_1,k_2,k_3,k_4)
		& = {\cal P}_{\zeta}^{-1}(k_1) {\cal P}_{\zeta}^{-1}(|\kk_1 + \kk_2|)  {\cal P}_{\zeta}^{-1}(k_4) \notag \\
		\times& \Bigg[ N_aN_{ab}N_{bc}N_c +  N_{ab}N_{bc}N_{cd}N_{da}\ln \left( \rm Min\{k_1,k_4,|\kk_1 + \kk_2| \right) 
		+ N_aN_{abc}N_{bcd}N_d \ln \left( |\kk_1 + \kk_2| L \right) \notag \\
		 & \ \ +  N_aN_{abc}N_{bd}N_{dc}\left(\ln \left( \rm Min\{k_4,|\kk_1 + \kk_2| \} L \right)
		   + \ln \left( \rm Min\{k_1,|\kk_1 + \kk_2| \} L \right) \right) \notag \\
		&  \ \ + N_{ab}N_{abc}N_{cd}N_d( \ln(k_1 L) + \ln(k_4 L) )  
		\Bigg]P_{\zeta}(k_1) P_{\zeta}(|\kk_1 + \kk_2|) P_{\zeta}(k_4) \notag \\
		 + & \mbox{ 11 perm.}  .
		\label{}
\end{align}

In order to identify the local form $\tau_{NL}$,
we impose the counter collinear limit,
\begin{equation}
k_{t_1} \equiv |\kk_1 + \kk_2| \ll k_{t_2} \equiv k_1 \sim k_2 \sim k_{t_3} \sim k_4,
		\label{}
\end{equation}
and we find, 
\begin{align}
		\tau_{NL}(k_{t_1},k_{t_2}) = &   {\cal P}_{\zeta}^{-1}(k_{t_1}) {\cal P}_{\zeta}^{-2}(k_{t_2})   
		\Big[ N_aN_{ab}N_{bc}N_c + 2 N_{ab}N_{abc}N_{cd}N_d \ln(k_{t_{2}} L)   \notag \\
		 & + \Big( N_{ab}N_{bc}N_{cd}N_{da} + N_aN_{abc}N_{bcd}N_d +  2 N_aN_{abc}N_{bd}N_{dc} \Big)\ln (k_{t_1} L )  \Big] ,
		\label{}
\end{align}
where the diagrams for $\tau_{NL}$ with 1-loop corrections are shown in Fig.~(\ref{fig:tau-1loop}).

These expressions for $f_{NL}$ and $\tau_{NL}$ in this Appendix are different from 
ones in \cite{Kawakami/etal:2009} and \cite{Suyama/etal}.

\section{Loop integral : all loop}
We give the power spectrum, bispectrum and trispectrum with all loop corrections from a diagrammatical method.
Each diagrams have coefficients and $\ln (kL)$ factors.
Therefore, the all loop expressions are given by
\begin{align}
		\sum_{\mbox {number of diagrams}}
		\left(
		\begin{array}{cc}
				\mbox{Coefficients }
		\end{array}
		\right)
		\times 
        \left(
		\begin{array}{cc}
				\mbox{Diagrams}
		\end{array}
		\right) 
		\times
		\left(
		\begin{array}{cc}
				\mbox{$\ln (kL)$ factors} 
		\end{array}
		\right)  ,
		\label{}
\end{align}
where the coefficients are given by
\begin{equation}
		\left(
		\begin{array}{cc}
				\mbox{coefficients of } \\ \mbox{the $\delta N$ expansion}
		\end{array}
		\right)
		\times 
        \left(
		\begin{array}{cc}
				\mbox{combinations of} \\ \mbox{the lines of propagators}
		\end{array}
		\right) 
		\times
		\left(
		\begin{array}{cc}
				\mbox{coefficients arising from} \\ \mbox{loop integrals} .
		\end{array}
		\right)  .
		\label{base}
\end{equation}

\subsection{Power spectrum}
The two-point function for $\zeta$ with all order is roughly expressed  by
\begin{equation}
		\left\langle \zeta \zeta \right \rangle
		= \sum_{n_1=1}^{\infty} \sum_{n_2=1}^{\infty}
		\frac{N_{a_{n_1}}}{n_1!}
		\frac{N_{a_{n_2}}}{n_2!} 
		\left\langle \delta \varphi^{a_{n_1}}_*  
		\delta \varphi^{a_{n_2}}_*  \right\rangle .
		\label{}
\end{equation}

\begin{figure}[t]
		\begin{center}
				\psfig{figure=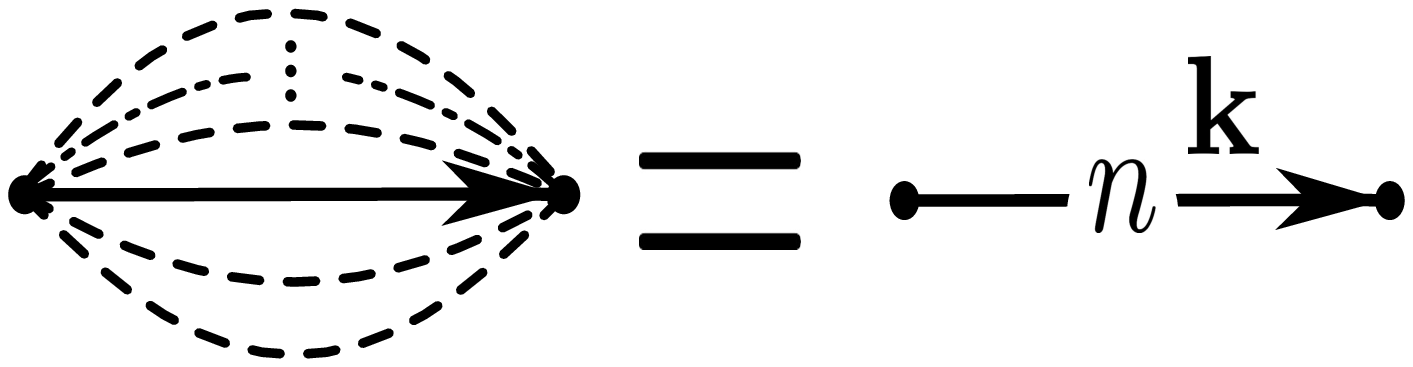}
		\end{center}
		\caption{The power spectrum including $n-1$ loops.
		        The RHS of this diagram denotes briefly that 
				the $n$ lines propagate momentum $\kk$.}
		\label{fig:power}
\end{figure}

Fig.~(\ref{fig:power}) means that
the two-point are connected by $n$ lines of the propagator
and these lines propagate the momentum $\kk$.
We first find that the number of propagators must be $n_1 = n_2 \equiv n$,
and the number of the combination to connect each lines is $n!$.
Therefore we find that Eq.~(\ref{base}) for power spectrum is
\begin{equation}
		\left( \frac{1}{n!} \right) \left( \frac{1}{n!} \right) 
		\times (n!) 
		\times
        \left(\begin{array}{cc}
				\mbox{coefficients arising from} \\ \mbox{loop integrals} .
		\end{array}\right) .
		\label{}
\end{equation}

Next, the loop integral for the power spectrum is given  by
\begin{equation}
		\int d^3p_1 \dots d^3p_n 
		\frac{1}{|\kk-\pp_1-\dots-\pp_n|^3} \frac{1}{p_1^3} \dots \frac{1}{p_n^3}  .
		\label{}
\end{equation}
We can evaluate this integral by using the {\it pole approximation}.
We assume that the main contribution of this integral come from around
each poles such as $\pp_1 = 0$, $\pp_2=0$, \dots, $\pp_{n-1} = 0$ and $\kk = \pp_1 + \dots + \pp_{n-1}$,
and they generate the factor $\ln (k L)$.
This means that
only one line chosen in $n$ lines propagates the momentum $\kk$,
and the other $n-1$ lines are integrated, generating the factor $\ln (k L)$.

Therefore the coefficient becomes
\begin{equation}
		\left( \frac{1}{n!} \right) \left( \frac{1}{n!} \right) 
		\times (n!) 
		\times n 
		= \frac{n}{n!}
		\label{}
\end{equation}
and the scale-dependence is given by $[\ln(k L)]^{n-1}$.

Finally, we can derive the expression of the power spectrum with all loop corrections:
\begin{equation}
		{\cal P}_{\zeta} = \sum_{n=1}^{\infty} \left( \frac{n}{n!}  \right) N_{a_n}N_{a_n} [\ln (kL)]^{n-1} ,
		\label{}
\end{equation}
where $N_{a_n} N_{a_n}$ is defined as
\begin{equation}
		N_{a_n} N_{a_n} \equiv 
		{\cal G}^{a_1 b_1}_*{\cal G}^{a_2 b_2}_* \dots {\cal G}^{a_n b_n}_*
		N_{a_1 a_2 \dots a_n} N_{b_1 b_2 \dots b_n} .
		\label{}
\end{equation}
\subsection{Bispectrum}
The three-point function for $\zeta$ with all order is roughly expressed  by
\begin{equation}
		\left\langle \zeta \zeta \zeta \right \rangle
		= \sum_{n_1=1}^{\infty} \sum_{n_2=1}^{\infty}\sum_{n_3=1}^{\infty}
		\frac{N_{a_{n_1}}}{n_1!}
		\frac{N_{a_{n_2}}}{n_2!} 
		\frac{N_{a_{n_3}}}{n_3!} 
		\left\langle \delta \varphi^{a_{n_1}}_*  
		\delta \varphi^{a_{n_2}}_* 	\delta \varphi^{a_{n_3}}_* \right\rangle .
		\label{}
\end{equation}

\begin{figure}[t]
		\begin{center}
				\psfig{figure=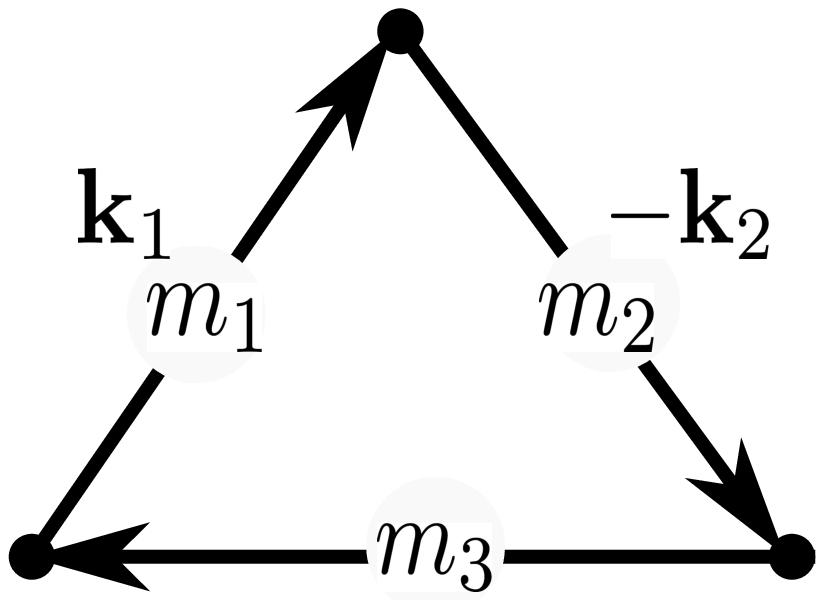}
		\end{center}
		\caption{Diagram for bispectrum including $(m_1 + m_2 + m_3 -2)$ loops. }
		\label{fig:bispectrum}
\end{figure}

Fig.~(\ref{fig:bispectrum})
means that 
each three points are connected by the $m_1$, $m_2$ and $m_3$ lines of propagators
and the two lines for $m_1$ and $m_2$ propagate the momenta momentum $\kk_1$ and $\kk_1 + \kk_3 = - \kk_2$.
For example, when $m_1 = m_2 = 1$ and $m_3=0$, the diagram means the tree level one.
First, we find that
the $m_1$, $m_2$ and $m_3$ lines is generated from each points $n_1$, $n_2$ and $n_3$ respectively and
the combinations to connect each lines are $m_1!$, $m_2!$ and $m_3!$ as like the case of the power spectrum.
Therefore, Eq.(\ref{base}) for the bispectrum is
\begin{align}
		\left( 	\frac{1}{n_1!} \frac{1}{n_2!} \frac{1}{n_3!} \right) 
	&	\times
		\left( \frac{n_1!}{m_1!m_2!} \frac{n_2!}{m_1!m_3!} \frac{n_3!}{m_2! m_3!} \right)
		\left( m_1!m_2!m_3! \right) 
		\times
        \left(\begin{array}{cc}
				\mbox{coefficients generated from} \\ \mbox{loop integrals} .
		\end{array}\right) \notag \\
		& = \frac{1}{m_1! m_2! m_3!} 
		\times 
		\left(
		\begin{array}{cc}
				\mbox{coefficients generated from} \\ \mbox{loop integrals}
		\end{array}
		\right) .
		\label{}
\end{align}

Next, we consider about the contributions of the loop integrals.
Without loss of generality,
we can consider that 
the all $m_3$ lines are integrated 
and generate the factor $\ln (\rm Min \{k_1,k_2 \})$.
On the other hand, the two lines in $m_1$ and $m_2$ 
propagate the momentum $\kk_1$, $-\kk_2$,
and the $m_1-1$ and $m_2-1$ lines are integrated, generating
the factor $\ln(k_1 L)$ and $\ln (k_2 L)$.
Then, the coefficient generated from the loop integral is $m_1 m_2$.
We apply the same discussion for the cases such as all $m_1$ and $m_2$ are integrated,
generating the coefficients of $m_2m_3$ and $m_1m_3$ respectively.
According to above discussion,
we find
\begin{align}
		B_{\zeta}(k_1,k_2,k_3) = & 
	  \sum_{m_1 = 0}^{\infty} \sum_{m_2=0}^{\infty} \sum_{m_3=0}^{\infty} 
      \frac{1}{ m_1! m_2! m_3!} N_{a_{m_1} b_{m_2}} N_{b_{m_2} c_{m_3}} N_{c_{m_3} a_{m_1}}  \notag \\
	& \times (2\pi^2)^2\Bigg[ m_1m_2\frac{[\ln{k_1 L}]^{m_1-1} [\ln{k_2 L}]^{m_2-1}[\ln {\rm Min}\{k_1,k_2\} L]^{m_3}}{k_1^3 k_2^3} \notag \\
	& \hspace{2cm} + m_1m_3\frac{[\ln{k_2 L}]^{m_3-1} [\ln{k_3 L}]^{m_1-1}[\ln {\rm Min}\{k_3,k_2\} L]^{m_2}}{k_3^3 k_2^3} \notag \\
	& \hspace{2cm} + m_3m_2\frac{[\ln{k_1 L}]^{m_3-1} [\ln{k_3 L}]^{m_2-1}[\ln {\rm Min}\{k_1,k_3\} L]^{m_1}}{k_1^3 k_3^3} \Bigg] \notag \\
	= &	  \sum_{m_1 = 1}^{\infty} \sum_{m_2=0}^{\infty} \sum_{m_3=1}^{\infty} 
		\frac{m_1m_3}{ m_1! m_2! m_3!} N_{a_{m_1} b_{m_2}} N_{b_{m_2} c_{m_3}} N_{c_{m_3} a_{m_1}}  \notag \\
	    &  \times  (2\pi^2)^2\Bigg[ \frac{[\ln{k_1 L}]^{m_3-1} [\ln{k_2 L}]^{m_1-1}[\ln {\rm Min}\{k_1,k_2\} L]^{m_2}}{k_1^3 k_2^3} \notag \\
	    &  \hspace{2cm}  + \frac{[\ln{k_2 L}]^{m_3-1} [\ln{k_3 L}]^{m_1-1}[\ln {\rm Min}\{k_3,k_2\} L]^{m_2}}{k_3^3 k_2^3} \notag \\
	    &  \hspace{2cm} + \frac{[\ln{k_1 L}]^{m_3-1} [\ln{k_3 L}]^{m_1-1}[\ln {\rm Min}\{k_1,k_3\} L]^{m_2}}{k_1^3 k_3^3} \Bigg] .
		\label{}
\end{align}

To make the bispecturm correspond the local form non-linear parameter $f_{NL}$,
we take the squeezed limit $k_{b_1} \equiv k_1 \ll k_2 \simeq k_3 \equiv k_{b_2}$,
and finally we find
\begin{align}
		 \frac{6}{5}f_{NL}(k_{b_1},k_{b_2})  = & 
		\frac{1}{ {\cal P}_{\zeta}(k_{b_1}) {\cal P}_{\zeta}(k_{b_2})} 
		 \sum_{m_1 = 1}^{\infty} \sum_{m_2=0}^{\infty} \sum_{m_3=1}^{\infty} 
			[\ln{k_{b_1} L}]^{m_2+m_3-1} [\ln{k_{b_2} L}]^{m_1-1} \notag \\
			& \hspace{3cm} \times \frac{m_1m_3}{ m_1! m_2! m_3!}N_{a_{m_1} b_{m_2}} N_{b_{m_2} c_{m_3}} N_{c_{m_3} a_{m_1}}  ,
		\label{}
\end{align}
where indeed the case of $m_1 = m_3 = 1$ and $m_2 = 0$ means the tree level.

\subsection{Trispectrum}
\begin{figure}[t]
		\begin{center}
				\psfig{figure=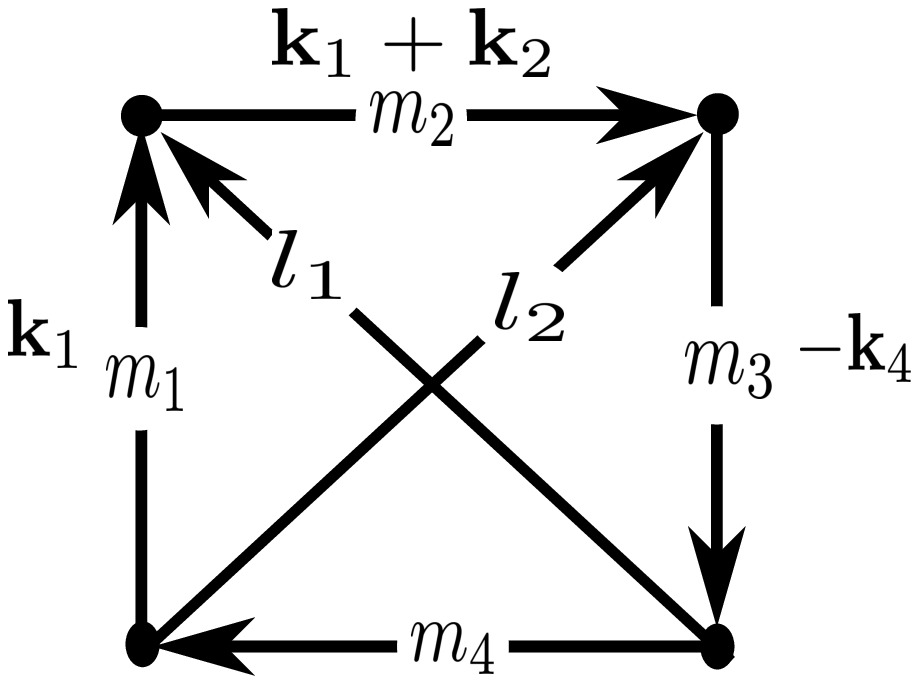}
		\end{center}
		\caption{Diagram of trispectrum corresponding to $\tau_{NL}$ 
		with $(m_1 + m_2 + m_3 + m_4 + l_1 + l_2 -3)$ loops.}
		\label{fig:Trispectrum}
\end{figure}
\begin{figure}[t]
		\begin{center}
				\psfig{figure=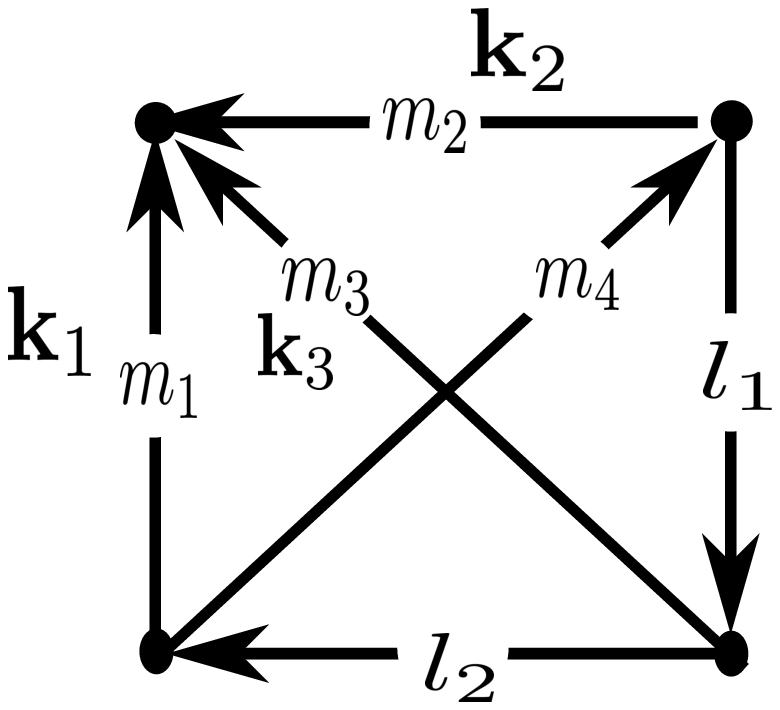}
		\end{center}
		\caption{Diagram of trispectrum corresponding to $g_{NL}$
		with $(m_1 + m_2 + m_3 + m_4 + l_1 + l_2 -3)$ loops.}
		\label{fig:Trispectrum2}
\end{figure}

In the case of the trispectrum,
we need to consider Fig.~(\ref{fig:Trispectrum}).
According to the similar discussion as like the case of the bispectrum,
we can find that Eq.~(\ref{base}) for trispectrum becomes
\begin{equation}
		\frac{1}{m_1!m_2!m_3!m_4!l_1!l_2!} \times 
		\left(
		\begin{array}{cc}
				\mbox{coefficients generated from} \\ \mbox{loop integrals}
		\end{array}
		\right) .
		\label{}
\end{equation}
For the coefficients arising from loop integrals,
there are twelve kinds of coefficients, that is,
we choose three from $m_1$, $m_2$, $m_3$, $m_4$, $l_1$, $l_2$ except for
$l_1m_1m_2$, $l_1m_3m_4$, $l_2m_2m_3$, $l_2m_1m_4$ which generate the other local form
corresponding $g_{NL}$.
However, we eventually can express the trispectrum only in terms of $m_1m_2m_3$ as like the case of the bispectrum,
and we find
\begin{align}
		& T_{\zeta}(k_1,k_2,k_3,k_4) \notag \\ 
		& = \sum_{m_1 = 1}^{\infty} \sum_{m_2=1}^{\infty}  \sum_{m_3=1}^{\infty} 
		 \sum_{m_4=0}^{\infty}  \sum_{l_1=0}^{\infty}  \sum_{l_2=0}^{\infty}  
         \frac{m_1m_2m_3}{m_1!m_2!m_3!m_4!l_1!l_2!}
		N_{d_{m_4}f_{l_2}a_{m_1}}N_{a_{m_1}e_{l_1}b_{m_2}}N_{b_{m_2}f_{l_2}c_{m_3}}N_{c_{m_3}e_{l_1} d_{m_4}}  \notag \\
		 &\hspace{0.5cm} \times (2\pi^2)^3 \Bigg\{\frac{1}{k_1^3|\kk_1 + \kk_2|^3 k_4^3}
		 \Big[ [\ln (k_1 L)]^{m_1-1} 	[\ln (|\kk_1 + \kk_2| L)]^{m_2-1} [\ln (k_4 L)]^{m_3-1}   \notag \\
		 & \hspace{2.5cm}  [\ln ( {\rm Min}\{|\kk_1 + \kk_2|,k_4 \}  L)]^{l_1}
	                    [\ln ( {\rm Min}\{k_1,|\kk_1 + \kk_2| \}   L)]^{l_2} 
                        [\ln ( {\rm Min}\{k_1,|\kk_1 + \kk_2|,k_4 \}  L)]^{m_4} \Big] \notag \\
		 & \hspace{13cm} + {\rm 11 perm.} \Bigg\} \notag 
		\label{}
\end{align}

When we take the collinear limit 
$k_{t_1} \equiv |\kk_1 + \kk_2| \ll k_{t_2} \equiv k_1 \sim k_2 \sim k_3 \sim k_4 $,
we finally find
\begin{align}
		\tau_{NL}(k_{t_1},k_{t_2})
		=& \frac{1}{ {\cal P}_{\zeta}(k_{t_1})  {\cal P}^2_{\zeta}(k_{t_2}) }
         \sum_{m_1 = 1}^{\infty} \sum_{m_2=1}^{\infty}  \sum_{m_3=1}^{\infty} 
		 \sum_{m_4=0}^{\infty}  \sum_{l_1=0}^{\infty}  \sum_{l_2=0}^{\infty}
		[\ln (k_{t_1}L)]^{m_2 + m_4+ l_1 + l_2-1} [\ln (k_{t_2} L)]^{m_1 + m_3-2} \notag \\
		&\hspace{2cm} \times\frac{m_1m_2m_3}{m_1!m_2!m_3!m_4!l_1!l_2!}  
		  N_{d_{m_4}f_{l_2}a_{m_1}}N_{a_{m_1}e_{l_1}b_{m_2}}N_{b_{m_2}f_{l_2}c_{m_3}}N_{c_{m_3}e_{l_1} d_{m_4}}  ,
		\label{}
\end{align}
where the case of $m_1 = m_2 = m_3 = 1$ and $m_4 = l_1 = l_2 = 0$ means
the tree level.

Similarly, we can derive the trispectrum corresponding to $g_{NL}$ from Fig.~(\ref{fig:Trispectrum2}).
\begin{align}
		& T_{\zeta}(k_1,k_2,k_3,k_4) \notag \\ 
		& = \sum_{m_1 = 1}^{\infty} \sum_{m_2=1}^{\infty}  \sum_{m_3=1}^{\infty} 
		 \sum_{m_4=0}^{\infty}  \sum_{l_1=0}^{\infty}  \sum_{l_2=0}^{\infty}  
         \frac{m_1m_2m_3}{m_1!m_2!m_3!m_4!l_1!l_2!}
		 N_{d_{l_2}f_{m_4}a_{m_1}}N_{a_{m_1}e_{m_3}b_{m_2}}N_{b_{m_2}f_{m_4}c_{l_1}}N_{c_{l_1}e_{m_3} d_{l_2}}   \notag \\
		 &\hspace{0.5cm} \times (2\pi^2)^3 \Bigg\{\frac{1}{k_1^3k_2^3k_3^3}
		 \Big[ [\ln (k_1 L)]^{m_1-1} 	[\ln (k_2 L)]^{m_2-1} [\ln (k_3 L)]^{m_3-1}   \notag \\
		 & \hspace{2.5cm} [\ln ( {\rm Min}\{k_1,k_2 \}   L)]^{m_4} 
                        [\ln ( {\rm Min}\{k_2,k_3 \}  L)]^{l_1}
                        [\ln ( {\rm Min}\{k_1,k_3 \}  L)]^{l_2} \Big] + {\rm 3 perm.} \Bigg\} \notag 
		\label{}
\end{align}
When we take the double squeezed limit
$k_{t_1} \equiv k_1 \simeq k_2 \ll k_3 \simeq k_4 \equiv k_{t_2}$,
we can derive $g_{NL}$ with all loop corrections as follows:
\begin{align}
		\frac{54}{25} g_{NL}(k_{t_1},k_{t_2})
		 = & \frac{1}{ {\cal P}^2_{\zeta}(k_{t_1})  {\cal P}_{\zeta}(k_{t_2}) }
         \sum_{m_1 = 1}^{\infty} \sum_{m_2=1}^{\infty}  \sum_{m_3=1}^{\infty} 
		 \sum_{m_4=0}^{\infty}  \sum_{l_1=0}^{\infty}  \sum_{l_2=0}^{\infty}  
		[\ln (k_{t_1}L)]^{m_1 + m_2 + m_4 + l_1 + l_2-2} [\ln (k_{t_2} L)]^{m_3-1} \notag \\
		&\hspace{2cm} \times \frac{m_1m_2m_3}{m_1!m_2!m_3!m_4!l_1!l_2!}
		 N_{d_{l_2}f_{m_4}a_{m_1}}N_{a_{m_1}e_{m_3}b_{m_2}}N_{b_{m_2}f_{m_4}c_{l_1}}N_{c_{l_1}e_{m_3} d_{l_2}}   \notag \\
		\label{}
\end{align}

\section{ Some calculations}
In this appendix, we show the following equations in Eq.~(\ref{C-S_inequality}),
\begin{equation}
	\tau_{NL}(k_L,k_S)  = 
	\frac{1}{ {\cal P}_{\zeta}(k_L) {\cal P}_{\zeta}^2(k_S) } \langle M(k_L.k_S), M(k_L.k_S) \rangle_G .
	\label{}
\end{equation}

When we focus only on $\langle M,M \rangle_G$,
we find 
\begin{align}
	     & \left \langle M(k_{L},k_{S}),M(k_{L},k_{S}) \right\rangle_G   \notag \\
         & = \sum_{\alpha=1}^{\infty}\sum_{m_1 = 1}^{\infty}\sum_{m_3 = 1}^{\infty}   \sum_{m_2=0}^{\alpha} \sum_{m_4=0}^{\alpha} 
		 [\ln{k_{L} L}]^{\alpha-1} [\ln{k_{S} L}]^{m_1+m_3-1} \notag \\
		 & \hspace{2cm}\times \frac{1}{4}\frac{m_1m_3 \alpha \alpha !}{ m_1!m_3! (\alpha-m_2)! (\alpha-m_4)! m_2! m_4!}  \notag \\
		 & \hspace{2cm}\times \frac{m_2! (\alpha-m_2)!}{\alpha!} \sum_{l=0}^{ M_1  }
		 \frac{m_4!}{ (M_2-l)! l! }  \frac{(\alpha-m_4)!}{ (\alpha-M_3-l)! (M_3-M_2+ l)! }  \notag \\
		 &\hspace{2cm}\times  N_{d_{\alpha-M_3-l}f_{l}a_{m_1}}N_{a_{m_1}e_{M_3-M_2+ l}b_{M_2-l}}
		 N_{b_{M_2-l}f_{l}c_{m_3}}N_{c_{m_3}e_{M_3-M_2+ l} d_{\alpha-M_3-l}} \notag \\
      & = \sum_{\alpha=1}^{\infty}\sum_{m_1 = 1}^{\infty}\sum_{m_3 = 1}^{\infty}   \sum_{m_2=0}^{\alpha} \sum_{m_4=0}^{\alpha}\sum_{l=0}^{ M_1  }
		 [\ln{k_{L} L}]^{\alpha-1} [\ln{k_{S} L}]^{m_1+m_3-1} \notag \\
		 & \hspace{2cm} \times \frac{1}{4}\frac{m_1m_3 \alpha }{ m_1!m_3! (M_2-l)! l! (\alpha-M_3-l)! (M_3-M_2+ l)!  } \notag \\
		 & \hspace{2cm} \times N_{d_{\alpha-M_3-l}f_{l}a_{m_1}}N_{a_{m_1}e_{M_3-M_2+ l}b_{M_2-l}}
		 N_{b_{M_2-l}f_{l}c_{m_3}}N_{c_{m_3}e_{M_3-M_2+ l} d_{\alpha-M_3-l}} ,
		 \label{MM}
\end{align}
where we define as
$M_1 \equiv {\rm Min}\{m_2,m_4,\alpha-m_2,\alpha-m_4  \}$, 
$M_2 \equiv {\rm Min}\{m_2,m_4\}$ and
$M_3 \equiv {\rm Max}\{m_2,m_4\}$.

In order to consider $M_1$, $M_2$ and $M_3$ specifically,
we decompose the range of $m_2$ and $m_4$ as follows:
(1) $\alpha = 2m_2$;
(2) $0 \le m_2 < \alpha-m_2$;
(3) $\alpha-m_2 < m_2 \le \alpha$.
The case (2) and (3) are equal when we replace $m_2$ by $\tilde{m}_2 \equiv \alpha - m_2$.
The same discussion for $m_4$ can be also applied.
Therefore, the summations for $m_2$ and $m_4$ are decomposed as
\begin{align}
		\sum_{m_2 = 0}^{\infty} \sum_{m_4=0}^{\infty}
		& = \left(\sum_{m_2=0}^{\infty} \delta_{\alpha = 2m_2} + 2\sum_{ \alpha/2 > m_2 \ge 0} \right) 
		\left( \sum_{m_4=0}^{\infty} \delta_{\alpha = 2m_4} + 2\sum_{\alpha/2 > m_4 \ge 0} \right)  \notag \\
		& = \sum_{m=0}^{\infty}  \delta_{\alpha = 2m}
		    + 4\sum_{m_4 > m_2}^{\infty} \sum_{m_2=0}^{\infty} \delta_{\alpha = 2m_4}
		    + 4\sum_{\alpha/2 > m \ge 0}
			+ 8 \sum_{\alpha/2 > m_4 > m_2 }\sum_{m_2 =0}^{\infty} ,
		\label{sum}
\end{align}
where $m$ is defined as $m\equiv m_2 = m_4$.

The first terms of Eq.~(\ref{sum}) means that
$M_1 = M_2 = M_3 = m = \alpha/2$.
Then the RHS of (\ref{MM}) is
\begin{align}
	    &  \sum_{m=0}^{\infty} \sum_{m_1 = 1}^{\infty}\sum_{m_3 = 1}^{\infty}  
		\sum_{l=0}^{ m }[\ln{k_{L} L}]^{2m-1}
         [\ln{k_{S} L}]^{m_1+m_3-1} \notag \\
		 & \hspace{2cm} \times \frac{1}{4}\frac{m_1m_3 (2m) }{ m_1!m_3! (m-l)!^2 l!^2 }N_{d_{m-l}f_{l}a_{m_1}}N_{a_{m_1}e_{l}b_{m-l}}
		        N_{b_{m-l}f_{l}c_{m_3}}N_{c_{m_3}e_{l} d_{m-l}} \notag \\
		 & = \sum_{\tilde{m}=0}^{\infty} \sum_{m_1 = 1}^{\infty}\sum_{m_3 = 1}^{\infty}  
		 \sum_{l=0}^{ \infty }[\ln{k_{L} L}]^{2(\tilde{m}+l)-1} [\ln{k_{S} L}]^{m_1+m_3-1} \notag \\
		 & \hspace{2cm} \times
		 \frac{1}{4}\frac{2m_1m_3 (\tilde{m} + l) }{ m_1!m_3! (\tilde{m}!)^2 (l!)^2 }	N_{d_{\tilde{m}}f_{l}a_{m_1}}N_{a_{m_1}e_{l}b_{\tilde{m}}}
		 N_{b_{\tilde{m}}f_{l}c_{m_3}}N_{c_{m_3}e_{l} d_{\tilde{m}}} ,
		 \label{1}
\end{align}
where we defined as $\tilde{m} \equiv m-l$.
This term is described in Fig.~(\ref{fig:Trispectrum})
as the case of $m_2 = m_4 = \tilde{m}$ and $l_1 = l_2 = l$.

The second term of Eq.~(\ref{sum}) means that
$M_1 = M_2= m_2$ and $M_3 = m_4 = \alpha/2$.
Then the RHS of (\ref{MM}) is
\begin{align}
		&  \sum_{m_4 > m_2}^{\infty}\sum_{m_1 = 1}^{\infty}\sum_{m_3 = 1}^{\infty}  \sum_{m_2=0}^{\infty}\sum_{l=0}^{ m_2  } 
		 [\ln{k_{L} L}]^{2m_4-1}
         [\ln{k_{S} L}]^{m_1+m_3-1} \notag \\
		 & \hspace{2cm}	\times \frac{2 m_1m_3 m_4 }{ m_1!m_3! (m_2-l)! l! (m_4-l)! (m_4-m_2+ l)! }	 \notag \\
		 & \hspace{2cm} \times N_{d_{m_4-l}f_{l}a_{m_1}}N_{a_{m_1}e_{m_4-m_2+ l}b_{m_2-l}}
		                       N_{b_{m_2-l}f_{l}c_{m_3}}N_{c_{m_3}e_{m_4-m_2+ l} d_{m_4-l}} \notag \\
	    & = \sum_{\tilde{m}_4>\tilde{m}_2}^{\infty}\sum_{m_1 = 1}^{\infty}
	        \sum_{m_3 = 1}^{\infty}\sum_{\tilde{m}_2=0}^{\infty}\sum_{l_1=0}^{ \infty } 
	        \sum_{l_2=0}^{\infty} \delta_{\tilde{m}_4-\tilde{m}_2 = l_2-l_1}
	   	    [\ln{k_{L} L}]^{\tilde{m}_2 + \tilde{m}_4 + l_1 + l_2 - 1} [\ln{k_{S} L}]^{m_1+m_3-1} \notag \\
		 &  \hspace{2cm} \times 
            \frac{ m_1m_3 (\tilde{m}_2 + \tilde{m}_4 + l_1 + l_2)}{ m_1!m_3! \tilde{m}_2! l_1! \tilde{m}_4! l_2!  }
		     N_{d_{\tilde{m}_4}f_{l_1}a_{m_1}}N_{a_{m_1}e_{l_2}b_{\tilde{m}_2}}
		     N_{b_{\tilde{m}_2}f_{l_1}c_{m_3}}N_{c_{m_3}e_{l_2} d_{\tilde{m}_4}}  ,
		 \label{2}
\end{align}
where we defined as 
$\tilde{m}_2 \equiv m_2-l $, $\tilde{m}_4 \equiv m_4-l$ and $l_2 \equiv \tilde{m}_4 - \tilde{m}_2 + l$ and  $l_1 \equiv l$.

The third term of Eq.~(\ref{sum}) means that
$M_1 = M_2 = M_3 = m$.
Then the RHS of (\ref{MM}) is
\begin{align}
		&  \sum_{\alpha>m}^{\infty}\sum_{m_1 = 1}^{\infty}\sum_{m_3 = 1}^{\infty}  
		\sum_{m=0}^\infty \sum_{l=0}^{ m  } [\ln{k_{L} L}]^{\alpha-1} [\ln{k_{S} L}]^{m_1+m_3-1} \notag \\
		 & \hspace{2cm} 	\times 	\frac{m_1m_3 \alpha }{ m_1!m_3! (m-l)! (\alpha-m-l)! (l!)^2  }
		 N_{d_{\alpha-m-l}f_{l}a_{m_1}}N_{a_{m_1}e_{l}b_{m-l}}
		 N_{b_{m-l}f_{l}c_{m_3}}N_{c_{m_3}e_{l} d_{\alpha-m-l}} \notag \\
		 & = \sum_{\tilde{m}_4 > \tilde{m}_2 }^{\infty}\sum_{m_1 = 1}^{\infty}\sum_{m_3 = 1}^{\infty}  
		 \sum_{\tilde{m}_2=0}^{\infty} \sum_{l=0}^{ \infty  } 
		 [\ln{k_{L} L}]^{\tilde{m}_2 + \tilde{m}_4 + 2l-1}[\ln{k_{S} L}]^{m_1+m_3-1} \notag \\
		 & \hspace{2cm} 	\times \frac{m_1m_3 (\tilde{m}_2 + \tilde{m}_4 + 2l) }{ m_1!m_3! \tilde{m}_2! \tilde{m}_4! (l!)^2  }
		 N_{d_{\tilde{m}_4}f_{l}a_{m_1}}N_{a_{m_1}e_{l}b_{\tilde{m}_2}}
		 N_{b_{\tilde{m}_2}f_{l}c_{m_3}}N_{c_{m_3}e_{l} d_{\tilde{m}_4}} \notag \\
		& = \sum_{\tilde{m}_4 > \tilde{m}_2 }^{\infty}\sum_{m_1 = 1}^{\infty}\sum_{m_3 = 1}^{\infty}  
		 \sum_{\tilde{m}_2=0}^{\infty} \sum_{l=0}^{ \infty  } 
		 [\ln{k_{L} L}]^{\tilde{m}_2 + \tilde{m}_4 + 2l-1}[\ln{k_{S} L}]^{m_1+m_3-1} \notag \\
		 & \hspace{2cm} 	\times \frac{1}{2} \frac{m_1m_3 (\tilde{m}_2 + \tilde{m}_4 + 2l) }{ m_1!m_3! \tilde{m}_2! \tilde{m}_4! (l!)^2  }
		 N_{d_{\tilde{m}_4}f_{l}a_{m_1}}N_{a_{m_1}e_{l}b_{\tilde{m}_2}}
		 N_{b_{\tilde{m}_2}f_{l}c_{m_3}}N_{c_{m_3}e_{l} d_{\tilde{m}_4}} \notag \\
		 & \hspace{1cm}  +  \sum_{\tilde{m}=0}^{\infty}\sum_{m_1 = 1}^{\infty}\sum_{m_3 = 1}^{\infty}  
		 \sum_{l_2 > l_1}^{\infty} \sum_{l_1=0}^{ \infty  } 
		 [\ln{k_{L} L}]^{2\tilde{m} + l_1 + l_2 - 1}[\ln{k_{S} L}]^{m_1+m_3-1} \notag \\
		 & \hspace{2cm} 	\times  \frac{1}{2} \frac{m_1m_3 (2\tilde{m} + l_1 + l_2) }{ m_1!m_3! (\tilde{m}!)^2 l_1! l_2!  }
		 N_{d_{\tilde{m}}f_{l_2}a_{m_1}}N_{a_{m_1}e_{l_1}b_{\tilde{m}}}
		 N_{b_{\tilde{m}}f_{l_2}c_{m_3}}N_{c_{m_3}e_{l_1}d_{\tilde{m}}},
		 \label{3}
\end{align}
where we defined as
$\tilde{m}_2 \equiv m -l$, $\tilde{m}_4 \equiv \alpha - m -l$
and $\tilde{m} \equiv \tilde{m}_2 = \tilde{m}_4$ and $l = l_1 = l_2$.
The last equal in Eq.~(\ref{3})
is shown by the fact that the replacement of $\tilde{m}_2 \leftrightarrow l_1$ and $\tilde{m_4} \leftrightarrow l_2$ 
does not the form of the equation.
In the explanation using Fig.~(\ref{fig:Trispectrum}),
this implies that the both cases of  $m_4 > m_2$, $l_2 = l_1 = l$ and $m_4 = m_2 = \tilde{m}$, $l_2 > l_1$ are equivalent.

Finally, the forth term of Eq.~(\ref{sum}) means that
$M_1 = M_2 = m_2$ and $M_3 = m_4$.
Then the RHS of (\ref{MM}) is
\begin{align}
		&  \sum_{\alpha > m_4}^{\infty}\sum_{m_1 = 1}^{\infty}\sum_{m_3 = 1}^{\infty} 
		\sum_{m_4 > m_2}^{\infty} \sum_{m_2 =0}^{\infty}  \sum_{l=0}^{ m_2  } 
		 [\ln{k_{L} L}]^{\alpha-1} [\ln{k_{S} L}]^{m_1+m_3-1} \notag \\
		 & \hspace{2cm}	\times 	\frac{2 m_1m_3 \alpha }{ m_1!m_3! (m_2-l)! l! (\alpha-m_4-l)! (m_4-m_2+ l)!  } \notag \\
		 & \hspace{2cm} \times 
		 N_{d_{\alpha-m_4-l}f_{l}a_{m_1}}N_{a_{m_1}e_{m_4-m_2+ l}b_{m_2-l}}
		 N_{b_{m_2-l}f_{l}c_{m_3}}N_{c_{m_3}e_{m_4-m_2+ l} d_{\alpha-m_4-l}}\notag \\
		&=  \sum_{(\tilde{m}_4- \tilde{m}_2) > (l_2-l_1)}^{\infty}  \sum_{m_1 = 1}^{\infty} \sum_{m_3 = 1}^{\infty} 
		   \sum_{\tilde{m}_2=0}^{\infty}   \sum_{l_2 > l_1}^{\infty} \sum_{l_1=0}^{\infty}   
		  	[\ln{k_{L} L}]^{\tilde{m}_2 + \tilde{m}_4 + l_2 + l_1-1} [\ln{k_{S} L}]^{m_1+m_3-2} \notag \\
		& \hspace{2cm} \times \frac{2 m_1m_3(\tilde{m}_2 + \tilde{m}_4 + l_2+l_1)}{m_1!m_3!\tilde{m}_2!\tilde{m}_4!l_1! l_2! } 
		N_{a_{m_1}f_{l_2}d_{\tilde{m}_4}}N_{a_{m_1}e_{l_1}b_{\tilde{m}_2}}N_{b_{\tilde{m}_2}f_{l_2}c_{m_3}}N_{c_{m_3}e_{l_1} d_{\tilde{m}_4}},
		\label{4}
\end{align}
where we defined as $\tilde{m}_2 \equiv m_2-l$,
$\tilde{m}_4 \equiv \alpha-m_4-l$, 
$l_2 \equiv m_4-m_2+l$ and $l_1\equiv l$.

Adding Eq.~(\ref{2}) and Eq.~(\ref{4}),
we find
\begin{align}
		 & \sum_{\tilde{m}_4 > \tilde{m}_2}^{\infty}  \sum_{m_1 = 1}^{\infty} \sum_{m_3 = 1}^{\infty} 
		   \sum_{\tilde{m}_2=0}^{\infty}   \sum_{l_2 > l_1}^{\infty} \sum_{l_1=0}^{\infty}   
		  	[\ln{k_{L} L}]^{\tilde{m}_2 + \tilde{m}_4 + l_2 + l_1-1} [\ln{k_{S} L}]^{m_1+m_3-2} \notag \\
		& \hspace{2cm} \times \frac{ m_1m_3(\tilde{m}_2 + \tilde{m}_4 + l_2+l_1)}{m_1!m_3!\tilde{m}_2!\tilde{m}_4!l_1! l_2! } 
		N_{a_{m_1}f_{l_2}d_{\tilde{m}_4}}N_{a_{m_1}e_{l_1}b_{\tilde{m}_2}}N_{b_{\tilde{m}_2}f_{l_2}c_{m_3}}N_{c_{m_3}e_{l_1} d_{\tilde{m}_4}},
		\label{5}
\end{align}
where this term means a diagram of $m_4 > m_2$ and $l_2 > l_1$ in Fig.~(\ref{fig:Trispectrum}).

Adding Eq.~(\ref{1}) and Eq.~(\ref{3}) and Eq.~(\ref{5}) ,
we finally can arrive at
\begin{align}
		&  \frac{1}{ {\cal P}_{\zeta}(k_{L}) {\cal P}^2_{\zeta}(k_{S})} 
           \left \langle M(k_{L},k_{S}),M(k_{L},k_{S}) \right\rangle_G  \notag \\
         & =  \frac{1}{ {\cal P}_{\zeta}(k_{L}) {\cal P}^2_{\zeta}(k_{S})} 
         \sum_{m_1 = 1}^{\infty}\sum_{m_3=1}^{\infty} \sum_{m_2=0}^{\infty} 
		 \sum_{m_4=0}^{\infty}  \sum_{l_1=0}^{\infty}  \sum_{l_2=0}^{\infty}
		 [\ln k_{L}L]^{m_2 + m_4+ l_1 + l_2-1} [\ln k_{S} L]^{m_1 + m_3-2} \notag \\
		 & \hspace{2cm} \times \frac{1}{4}\frac{m_1m_3(m_2 + m_4 + l_1 + l_2)}{m_1!m_2!m_3!m_4!l_1!l_2!}  
		  N_{a_{m_1}f_{l_2}d_{m_4}}N_{a_{m_1}e_{l_1}b_{m_2}}N_{b_{m_2}f_{l_2}c_{m_3}}N_{c_{m_3}e_{l_1} d_{m_4}} \notag \\
         & = \frac{1}{ {\cal P}_{\zeta}(k_{L}) {\cal P}^2_{\zeta}(k_{S})} 
         \sum_{m_1 = 1}^{\infty}\sum_{m_2=1}^{\infty} 
		 \sum_{m_3=1}^{\infty} \sum_{m_4=0}^{\infty}  \sum_{l_1=0}^{\infty}  \sum_{l_2=0}^{\infty}
		 [\ln k_{L}L]^{m_2 + m_4+ l_1 + l_2-1} [\ln k_{S} L]^{m_1 + m_3-2} \notag \\
		 & \hspace{3cm} \times \frac{m_1m_2m_3}{m_1!m_2!m_3!m_4!l_1!l_2!}  
		 N_{a_{m_1}f_{l_2}d_{m_4}}N_{a_{m_1}e_{l_1}b_{m_2}}N_{b_{m_2}f_{l_2}c_{m_3}}N_{c_{m_3}e_{l_1} d_{m_4}} \notag \\
		 & = \tau({k_L,k_S} ) .
		\label{}
\end{align}

\bibliographystyle{JHEP}
\bibliography{sugiyama_sixth}

\end{document}